\documentclass[12pt]{article}
\usepackage{amssymb}
\usepackage{graphicx}
\usepackage{epsf}
\usepackage{lscape}
\usepackage{epstopdf}
\usepackage[cp866]{inputenc}
\usepackage[T2A]{fontenc}
\usepackage[english]{babel}
\usepackage{amsmath}
\begin{document}

\title{\large{\bf Dispersion (asymptotic) theory of charged particle transfer reactions at low energies and nuclear astrophysics: II. the asymptotic normalization coefficients and their nuclear-astrophysical application}}
\author{    R. Yarmukhamedov,$^{\rm{1,2}}$\thanks{Corresponding author, E-mail:
rakhim@inp.uz}\, K. I. Tursunmakhatov$^{\rm{3}}$ and N. Burtebayev$^{\rm{2,\,4}}$   }

 \maketitle{\it
$^{\rm{1}}$Institute of Nuclear Physics, 100214 Tashkent, Uzbekistan\\
$^{\rm{2}}$Al Farabi Kazakh National University, 050040 Almaty, Kazakhstan\\
$^{\rm{3}}$  Gulistan State University,  120100 Gulistan, Uzbekistan\\
$^{\rm{4}}$ Institute of Nuclear Physics, 050032 Almaty,  Kazakhstan\\
}

\bigskip

\begin{abstract}
 Within the asymptotic theory  proposed by authors R. Yarmukhamedov and K.I. Tursunmakhatov [Phys. Rev. C (submitted 2019)] for the peripheral sub- and above-barrier transfer $A$($x$, $y$)$B$  reaction in the three-body ($A$, $a$ and $y$) model ($ x$= $y$ + $a$ and $B$= $A$ + $a$, and $ a$  is a transferred particle),
 the analysis of the experimental angular distributions of the differential cross sections  is performed for the peripheral proton and triton transfer ${\rm{^9Be(^{10}B,\,^9Be)^{10}B}}$, ${\rm{^{16}O(^3He}},\,d{\rm{)^{17}F}}$ and ${\rm{^{19}F(}}p,\,\alpha{\rm{)^{16}O}}$ reactions at above-and sub-barrier projectile energies, respectively.  New estimates and their uncertainties are obtained for magnitudes of the  asymptotic normalization coefficients (respective the nuclear vertex constants) for ${\rm{^9Be}} \,+p \to\,{\rm{^{10}B}}$,   ${\rm{^{16}O}} \,+\,p \to\, {\rm{^{17}F}}$ and ${\rm{^{16}O}} \,+\,t\to{\rm{^{19}F}}$.  They are applied  for calculations of   the     astrophysical  $S$ factors   for the nuclear-astophysical ${\rm{^9Be(}}p,\,\gamma{\rm{)^{10}B}} $,    ${\rm{^{16}O(}}p,\,\gamma{\rm{)^{17}F}}$ and
 ${\rm{^{19}F(}}p,\,\alpha{\rm{)^{16}O)}}$ reactions at thermonuclear energies.  New values and their uncertainties   are obtained for the astrophysical $S$ factors  at stellar energies.
\end{abstract}

\newpage

 \vspace{1.0cm}
\begin{center}
 {\bf I. INTRODUCTION}
  \end{center}

\bigskip

  Results from earlier nuclear physics research at very level in  a hot Big Bang nucleosynthesis, which were predicted by astrophysicists within the framework  of  the standard cosmology,   let   to
 suspicion  that   spallation  reactions play the main role in   production  of light elements such as $ {\rm{B}}$, $ {\rm{O}}$, ${\rm{ F}}$ and etc.  See Refs.  \cite{Fowler1984,Rolf88}.
 Therefore, a reliable estimation of rates of
 different nuclear astrophysical processes responsible for the light element
abundance  is one of the most actual
problems of the modern nuclear astrophysics \cite{Adel2011}.
Solution of this problem is in turn impossible without obtaining  rather low
energy cross sections $\sigma(E)$ (or respective  astrophysical $S$ factors
$S(E)$) for such reactions.

Despite the impressive improvements in our understanding of such 
 reactions  made in the past decades (see Refs
[3--6] for example) ambiguities connected
 both with the extrapolation of the measured cross sections for some   specific nuclear-astrophysical
reactions above within  the stellar energy region and with the theoretical predictions for $\sigma
(E)$ (or $S(E)$) still exist. They may considerably influence  predictions of the
 standard solar model \cite{Fowler1984,Rolf88}.
  As a specific example below we consider the nowaday situation concerning
 the nuclear-astrophysical  $ {\rm{^9Be(}}p,\,\gamma{\rm{)^{10}B}} $, $ {\rm{^{16}O(}}p,\,\gamma{\rm{)^{17}F}} $ and $ {\rm{^{19}F(}}p,\,\alpha{\rm{)^{16}O}} $ reactions since the calculations of the corresponding astrophysical $S$ factors,  performed within
different methods [7--15], show noticeable spread exceeding the experimental errors.

The  $ {\rm{^9Be(}}p,\,\gamma{\rm{)^{10}B}} $ reaction plays an important role as one of the critical links   in  primordial and stellar nucleosynthesis of light elements in the $p$ shell \cite{Fowler1984,Rolf88,Ceciletal1992,Zah1995}. In \cite{Ceciletal1992}, the experimental astrophysical  $S$ factors ($S^{{\rm{exp}}}_{{\rm{1\,9}}}{\rm{(}}E{\rm{)}}$) for this reaction  were measured over the energy range 68$<E<$125 keV via the measurement of the branching ratio for the $ {\rm{^9Be(}}p,\,\gamma{\rm{)^{10}B}} $ and $ {\rm{^9Be(}}p,\,\alpha{\rm{)^6Li}} $ reactions (here and everywhere below,   the lower indexes in the astrophysical $S$ factor denote mass numbers of the colliding particles).  It was   revealed that  the $S^{{\rm{exp}}}_{{\rm{1\,9}}}{\rm{(}}E{\rm{)}}$ measured is   practically independent from the energy $E$. In \cite{Ceciletal1992}, these data    were then analyzed  within the framework of the two-body potential method  under two the assumptions  that the pure direct capture contributes  and  the spectroscopic factor for the ${\rm{^{10}B}}$  nucleus in the   (${\rm{^9Be}}+p$) configuration can be set equal to unity.   On the other hand,   the experimental energy dependence of the $S^{{\rm{exp}}}_{{\rm{1\,9}}}{\rm{(}}E{\rm{)}}$, measured  by authors of Ref.  \cite{Zah1995} over     the range most important to nuclear astrophysics (66$<E<$1620 keV), includes contributions from four the resonances and the direct capture as well as     their allowed  interference with each other.
In \cite{Wulf1998}, the $S^{{\rm{exp}}}_{{\rm{1\,9}}}{\rm{(}}E{\rm{)}}$  data of Ref. \cite{Zah1995}  were  analyzed within the two-body potential method  where the direct component ($S^{{\rm{DC}}}_{{\rm{1\,9}}}{\rm{(}}E{\rm{)}}$) of the calculated total astrophysical $S$ factor ($S_{{\rm{1\,9}}}{\rm{(}}E{\rm{)}}$) is presented as the product of   the spectroscopic factor mentioned above, which is  taken from \cite{Rolfs1973},  and the bilinear polynomial over the energy $E$.  The obtained result shows that the calculated $S_{{\rm{1\,9}}}{\rm{(}}E{\rm{)}}$ values are larger than the results of  \cite{Ceciletal1992}  by a factor of 4.2.
In \cite{Satt1999}, the analysis of the experimental data  \cite{Zah1995} has been performed within the modified $R$-matrix method in which the contribution from  the direct capture amplitude to  the total $S_{{\rm{1\,9}}}{\rm{(}}E{\rm{)}}$  is calculated using the  ``indirect measured'' ANC   for ${\rm{^9Be}}+p\to{\rm{^{10}B}}$   derived in Ref. \cite{Mukh2}  for   the ground and first three excited states of ${\rm{^{10}B}}$. In this connection, one notes that,  in \cite{Mukh2}, the ANC's above were obtained from the analysis of the precisely measured differential cross sections (DCS's) for the proton exchange ${\rm{^{10}B}}$-${\rm{^9Be}}$ reaction, which was     within the ``post'' form of the modified  
distorted-wave Born approximation (MDWBA). Hence,    the contribution of the three-body Coulomb effects in the full transition operator of the three-body DWBA amplitude is  taken into account in the first order of the pertubation theory over the Coulomb polarization potential $\Delta V^{{\rm{C}}}_{i,f}$ \cite{Mukh10}. While, as   shown in Ref. \cite{YTur2019}, when the residual ${\rm{^{10}B}}$ nucleus in the peripheral transfer reaction is formed especially  in excited bound states, this restriction over  $\Delta V^{{\rm {C}}}_{i,f}$ in  the transition operator does not guarantee the necessary accuracy   of the ``indirect measured'' ANC values for their  astrophysical application. Besides, in  \cite{Satt1999},  the channel  contribution \cite{Holt1978} to   the resonance $\gamma$-ray width in  the resonance component of the total $R$-matrix amplitude as a factor,   was ingored. Correct taking into account  this contribution to the resonance $\gamma$-ray width may influence the energy dependence of the resonance amplitude given by Eqs. (5)--(7) in \cite{Satt1999}. Therefore,  it would be  highly encouraged  an examination of a degree of a reliability  of    the   assumptions used in Refs. \cite{Satt1999,Mukh2}.

If the subsequent hydrogen burning of  ${\rm {^{19}F}}$ proceeds
 predominantly through the ${\rm {^{19}F(}}p,\,\alpha{\rm{)^{16}O}}$ reaction,  the ${\rm {^{16}O}}(p,\gamma){\rm {^{17}F}}$  reaction is the first one in a  link of the sequence of a four branch in CNO hydrogen burning proceeding  via ${\rm{^{16}O(}}p,\gamma{\rm {)^{17}F(}}e^+\nu_e$)${\rm {^{17}O(}}p,\gamma{\rm {)^{18}F(}}e^+\nu_e$)${\rm {^{18}O(}}p,\gamma{\rm {)^{19}F(}}p,\,\alpha{\rm{)^{16}O}}$. This changeover  from    the $pp$-chain to   the CNO cycle is observed  near $T_{{\rm{6}}}\approx$20 K (the Gamow's energy $E_G\approx$ 35.2 keV) \cite{Rolf88}. The ${\rm {^{16}O}}(p,\gamma){\rm {^{17}F}}$ reaction rate sensitively influences the ${\rm{^{17}O}}/{\rm{^{16}O}}$ isotopic ratio predicted  by models of massive AGB stars, where proton capture occurs at the base of the convective envelop. In second-generation   stars, whose stellar temperatures are higher than  those for the quiescent  CNO cycle,   the  ${\rm {^{19}F}}(p,\gamma){\rm {^{20}Ne}}$ and ${\rm {^{19}F(}}p,\,\alpha{\rm{)^{16}O}}$ reactions
    compete with each other in the hydrogen burning phase corresponding to the transition from  the hot CNO cycle to  the NeNa one.
Furthermore, the ${\rm {^{19}F(}}p,\,\alpha{\rm{)^{16}O}}$ reaction can play an important role both in hydogen-rich environment and in the AGB stars as the main sites of fluorine production \cite{Rolf88}.
  Despite its importance, the astrophysical $S$ factors have  still large uncertainties at astrophysical energies \cite{An99}. As it is seen from above,  exact knowledge of the rates of the  ${\rm {^{16}O}}(p,\gamma){\rm {^{17}F}}$ and ${\rm {^{19}F(}}p,\,\alpha{\rm{)^{16}O}}$ reactions is of great importance for modeling of nucleosynthesis in the hydrogen-burning massive stars.

   There are old  measurements  of the  ${\rm {^{16}O}}(p,\gamma){\rm {^{17}F}}$ reaction  close to the energy region of astrophysical interest (see \cite{Rolfs1973} and  references therein). However, these experiments did not distinguish between transitions into
the  ground ($E^*$=0.0; $J^{\pi}$=5/2$^+$) state and the   first  excited ($E^*$=0.497 MeV; $J^{\pi}$=1/2$^+$) state of  the residual ${\rm {^{17}F}}$ nucleus. In  \cite{Morl1997}, the experimental astrophysical $S$ factor, $S_{1\,16}^{{\rm{exp}}}(E)$, has been measured in the  energy range  200$\le E\le$3750 keV with  the separated ground and first excited states of  the residual ${\rm {^{17}F}}$ nucleus. They were then analyzed using the Woods-Saxon potential in the standard  two-body method under the assumption that the spectroscopic factors  for ${\rm {^{17}F}}$ in the  (${\rm {^{16}O}}+p$) configuration   can   be set to unity both for the ground  state and for the first excited one of ${\rm {^{17}F}}$. However, as  shown   in \cite{Art2009},  in reality,      there   are infinite  number of the  phase-equivalent Woods-Saxon potentials resulting    in the theoretical uncertainty about 50\%
  in     the calculated  $S_{1\,16}(E)$ values at stellar energies. Though, all these potentials lead to   the calculated  phase shifts for the $p{\rm {^{16}O}}$-scattering, which  are   in a well agreement with the experimental data within    the uncertainty up to $\sim$10\%.
 It follows from here that the results of Ref.  \cite{Morl1997} derived for    $S_{1\,16}(E)$   are strongly  model-dependent. It is mainly associated with the fact that  the spectroscopic factors above   cannot be determined unambiguously \cite{Art2009} and, so,  their values  should not be set to unity, a priori.  The  astrophysical $S$ factor  $S_{1\,16}(E)$ at stellar energies  was also calculated in \cite{Mukh1999} within the standard  two-body potential method using the ANC values for      ${\rm {^{16}O}}+p\to{\rm {^{17}F}}$(g.s) and  ${\rm {^{16}O}}+p\to{\rm {^{17}F}}$(0.497 MeV). They were derived in \cite{Mukh1999} from the analysis of the experimental DCS's of the peripheral proton transfer  ${\rm {^{16}O(^3He}},\,d {\rm {)^{17}F}}$ reaction, which was  performed within the post form of  the  modified DWBA.  As it was mentioned above, the post form of  the  modified DWBA cannot provide  the necessary accuracy  in  the ANC values  for their astrophysical application, especially, for the very weakly bound first excited state of ${\rm{^{17}F}}$.
This point  relates to  the ANC values obtained in Ref. \cite{Ar96} from the analysis of the same peripheral  proton transfer   reaction  in the finite-range  of the  ``post''-approximation of MDWBA.
  That apparently is one of the reasons why the central value for the square of the ANC for  ${\rm {^{16}O}}+p\to{\rm {^{17}F}}$(0.497 MeV) recommended in Refs. \cite{Mukh1999}  and \cite{Ar96} is  about 3.5$\sigma$ larger and  2.6$\sigma$ lesser, respectively,  than that recommended  in Refs. \cite{Art2009,YarBaye2011}. See Table \ref{table1} below.

   For the ${\rm {^{19}F(}}p,\,\alpha{\rm{)^{16}O}}$ reaction there are
  unpublished experimental data  by Lorenz-Wirzba \cite{HL1978}  and the data measured by authors of Refs. \cite{Ivano2015} at the  lowest sub-barrier
  proton projectile energies, including  nonresonant values ($\lesssim$ 500 keV).   The experimental angular distributions of the  differential cross sections of  Ref. \cite{HL1978} measured at  the proton projectile  energies   250, 350 and 450 keV have been quoted in Ref.  \cite{HAS1991}, where the   analysis has been performed  within the zero-range  approximation of the conventional  DWBA. As a consequence, it was revealed that  the ground state transition may be dominated by  a direct  mechanism  in the nonresonant energy region below the Coulomb barrier, involving the vicinity of    the AGB Gamow window ($\simeq$27--94 keV at $T_{{\rm{6}}}\simeq $40 K). Nevertheless, it occurs  the discrepancy between the absolute values  of the experimental  angular distributions  of Refs. \cite{HL1978} and \cite{Ivano2015} at rather close nonresonant proton projectile energies  by a factor of about 2. Besides,
the results of  the investigation of the ${\rm {^{19}F(}}p,\,\alpha{\rm{)^{16}O}}$ reaction at energies below the Coulomb barrier reported by different authors (see recent work of Ref. \cite{Ivano2015} and references therein) show a presence of    rather large spread  in the calculated values of  the  astrophysical $S$ factors at center-of-mass energies down 200 keV.   Therefore, the application of  the asymptotic theory  developed in Ref. \cite{YTur2019} for the sub-barrier ${\rm {^{19}F(}}p,\,\alpha{\rm{)^{16}O}}$ reaction  allows  to obtain new quantitative information both  about the direct mechanism at nonresonant projectile energies  and about  the possibility of extraction of the ``indirect determined'' ANC for ${\rm{^{16}O}}+t\to {\rm{^{19}F}}$.

  Below, we present the results of the analysis of the experimental angular distributions of the DCS's for the mentioned above   peripheral proton and triton transfer
 reactions \cite{Mukh2,Mukh1999,HL1978,Ivano2015}  and  their application for obtaining  a new information about the extrapolated astrophysical $S$ factors at stellar energies for   the corresponding specific
 nuclear-astrophysical processes.  The analysis will be performed  within framework of the asymptotic theory  developed in Ref. \cite{YTur2019} since all these reactions   are   related to  the  ``non-dramatic'' case. As noted in Refs. \cite{Mukh10,YTur2019},  the latter occurs  when the values of the Coulomb parameters for the two-body bound state wave functions  in the entrance and exit channels or their sum are not in the vicinity of  a  natural number.

\bigskip
{\bf VI.  ANALYSIS  OF   THE PERIPHERAL SUB- AND ABOVE-BARRIER TRITON AND PROTON REACTIONS }
  \bigskip

  In this section, we present the results of comparison of the calculated DCS's  with experimental data for the following  peripheral   proton and triton transfer  reactions:\\
   (I) ${\rm {^9Be(^{10}B,^9Be)^{10}B}}$  at the ${\rm{^{10}B}}$ incident  energy $E_{\rm{^{10}B}}$= 100 MeV \cite{Mukh2},\\
     (II)  ${\rm{^{16}O(^3He}}, d{\rm {)^{17}F}}$  at  $E_{\rm{^3He }}$= 29.75 MeV \cite{Mukh1999},\\
     (III) ${\rm{^{19}F(}}p,\,\alpha {\rm {)^{16}O}}$ \cite{Ivano2015,HL1978}   at six  sub-barrier proton projectile  energies.

The experimental angular distributions of the DCS's of the reaction (I)  are analyzed for the residual  ${\rm{^{10}B}}$ nucleus  populating the ground ($E^*$=0.0; $J^{\pi}$=3$^+$) state, the  first
    ($E^*$= 0.718 MeV; $J^{\pi}$=1$^+$), second ($E^*$= 1.740 MeV; $J^{\pi}$=0$^+$) ana third ($E^*$= 2.154 MeV; $J^{\pi}$=1$^+$) excited states (denoted by ${\rm{^{10}B_0}}$, ${\rm{^{10}B_1}}$, ${\rm{^{10}B_2}}$ and ${\rm{^{10}B_3}}$, respectively, below). The residual  ${\rm {^{17}F}}$ nucleus in the reaction (II) is formed in  the ground ($E^*$=0.0; $J^{\pi}$=$\frac{{\rm{5}}}{{\rm{2}}}^+$)   and   first ($E^*$=0.495 MeV; $\frac{{\rm{1}}}{{\rm{2}}}^+$) excited states (denoted by ${\rm {^{17}F_0}}$ and ${\rm {^{17}F_1}}$, respectively, below). For the reaction (III) populating the ground   state of the residual ${\rm {^{16}O}}$ nucleus,  two  the  sets of the independently measured  experimental data are considered, which correspond to the projectile proton energy ($E_p$=  250; 350 and 450 keV \cite{HL1978} (denoted by EXP-1978 below) and  $E_p$=  327; 387 and 486 keV \cite{Ivano2015} (denoted by EXP-2015 below).

 For the reactions considered above,   the orbital  $l_B$ and $l_x$  angular momentums of the transfer $a$  particle ($a$ is either proton or triton) in the bound  $B$ and $x$ nuclei ($B$ is either ${\rm{^{10}B}}_i$ or ${\rm{^{17}F}}_i$ or ${\rm{^{19}F}}$ and $x$ is either ${\rm^{10}}B_{{\rm{0}}}$ or ${\rm{^3He}}$ or $\alpha$ particle), respectively,  are  taken   equal to  $l_{{\rm{^{10}B}}_i}$= 1 ($i$=0--3),  $l_{{\rm{^{17}F_0}}}$= 2 and $l_{{\rm{^{17}F_1}}}$= $l_{{\rm{^{19}F}}}$= 0,  and $l_{{\rm{^3He}}}$= $l_{\alpha}$= 0.
  Since the energy of incident ${\rm{^3He}}$  in the reaction (II) is moderate, the contribution of the   $d$-state of  the ${\rm{^3He}}$ nucleus  in the vertex  ${\rm{^3He}}\to d\,+p$ is  neglectable small \cite{Blok77}.
   In this case, the total angular  $j_B$ and $j_x$ momentums, where $\mathbf{j}_{B}=\mathbf{l}_B+\mathbf{J}_a$ and $\mathbf{j}_{x}=\mathbf{l}_x+\mathbf{J}_a$ in which $\mathbf{J}_a$ is the spin of the transferred $a$ particle, are taken equal to $j_{{\rm{^{10}B_0}}}$=$j_{{\rm{^{10}B_2}}}$=  3/2, $j_{{\rm{^{17}F_0}}}$= 5/2, $j_{{\rm{^{19}F}}}$=1/2,
     and $j_{{\rm{^{17}F_1}}}$= $j_{\alpha}$= $j_{{\rm{^3He}}}$= 1/2,
  whereas $j_{{\rm{^{10}B_1}}}$= $j_{{\rm{^{10}B_3}}}$= 1/2 and 3/2.

 In this case, at the  fixed  values of the angular ($l_x$ and $l_B$) and total ($j_x$) orbital momentums defined above,
 the expression (51),   derived in  Ref.  \cite{YTur2019} for the  DCS of  the peripheral transfer $A$($x$, $y$)$B$ reaction (where $B=A+a$ and $x=y+a$, and $a$ is the transferred particle),   can be presented in the form
    \begin{equation}
\label{FBCS2}
\frac{d\sigma}{d\Omega}=C^{{\rm{2}}}_{x}\sum_{j_B}C^{{\rm{2}}}_{B;j_B}\tilde{\sigma}_{r_{{\rm{0}}}}{{\rm(}}j_B;E_i,\,\theta{\rm{)}},
\end{equation}
 where $C_{B;j_B}=C_{Aa;\,l_B\,j_B}$ and $C_{x}=C_{ay;\,l_x\,j_x}$ are the ANC's for $A+a\to B$ and $y+a\to x$ \cite{Blok77}, respectively, which are related to respective the nuclear vertex constants (NVCs)  for the virtual decays $B\to A+a$ and $x\to y+a$ by the simple relations given in \cite{Blok77} (see Eq. (7) in \cite{YTur2019} also);
  the $\tilde{\sigma}_{r_{{\rm{0}}}}{\rm{(}}j_B;E_i,\,\theta{\rm{)}}$ is a known function of  the center-of-mass scattering angle $\theta$ and  energy $E_i$  at fixed values of  the  quantum  ($l_x,\, j_x,\, l_B$) numbers above and the cut-off channel  $R_i^{{\rm{ch}}}$ and  $ R_f^{{\rm{ch}}}$ radii corresponding respectively  to   the entrance and exit channels.  They enter the lower limits of   the   radial integral of  the matrix element of the reaction  and can be  determined   by $R_i^{{\rm{ch}}}$=$r_{{\rm{0}}}$($A^{{\rm{1/3}}}+x^{{\rm{1/3}}}$) and  $ R_f^{{\rm{ch}}}$=$r_{{\rm{0}}}$($B^{{\rm{1/3}}}+y^{{\rm{1/3}}}$), where $r_{{\rm{0}}}$ is the nuclear interaction radius   and $D$ is a mass number of $D$ nucleus ($D=A, \,x, \,B$ and $y$). The optical potentials for the entrance and exit  channels were taken from Refs. \cite{Mukh2,Mukh1999} (the sets 1 and 2) and \cite{HAS1991}. Calculations were performed     using  the expression (51) of \cite{YTur2019}  and Eq. (\ref{FBCS2}) in which   the influence of the three-body ($A$, $a$ and $y$) Coulomb dynamics of  the transfer mechanism  on the peripheral partial amplitudes at $l_i>>$1 and $l_f>>$1 of the  reaction amplitude
is taken into account in a correct manner. As shown in   \cite{YTur2019}, this influence on      the  amplitudes of the  reactions  considered above    is also noticeable at least in the angular range of the main peak of the angular distributions. It  is also  related to the calculated reduced $\tilde{\sigma}_{r_{{\rm 0}}}(j_B;E_i,\,\theta)$  cross sections.   The      values   of the product of the square of the ANC's above  and $r_{{\rm 0}}$  entering Eq. (\ref{FBCS2}) and their uncertainties  giving the best fit to the experimental DCS's   within the experimental errors have been defined by minimizing the quantity   $\chi^{{\rm 2}}$ in the fitted data only in the angular region of the main peak of the angular distribution.

\bigskip

   {\bf A. Asymptotic normalization coefficients for ${\rm {^9Be}}+p\to {\rm{^{10}B}}$,  ${\rm {^{16}O}}+p\to{\rm{^{17}F}}$ and   $ {\rm {^{16}O}}+t\to{\rm{^{19}F}}$ }

\bigskip

\begin{figure}
\begin{center}
\includegraphics[height=15cm, width=12cm]{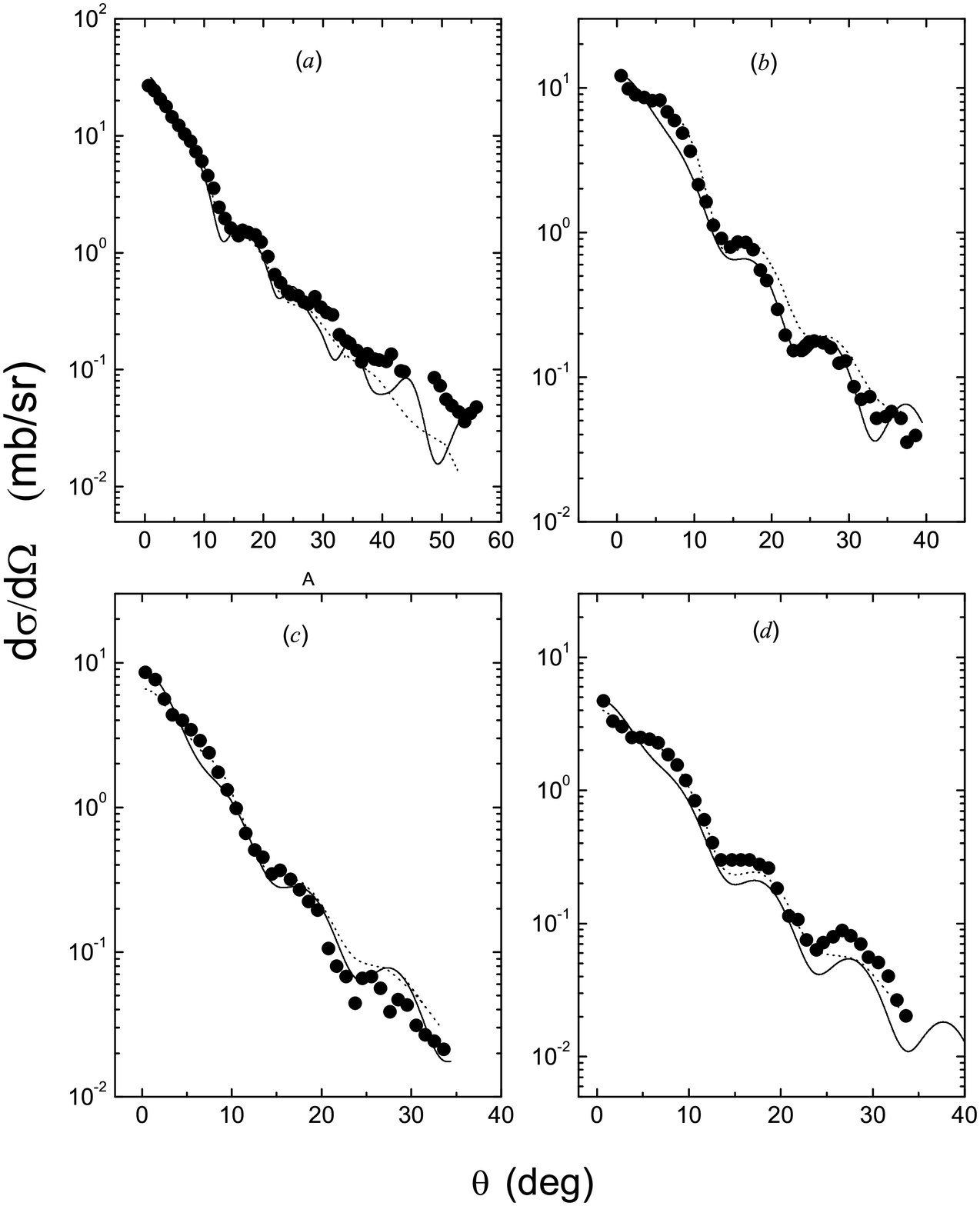}
 \end{center}
\caption{\label{fig1} The differential cross sections   for the ${\rm {^9Be(^{10}B}},\,{\rm{^9Be)^{10}B}}$    reaction at $E_{\rm{^{10}B}}$= 100 MeV.
The points are the experimental data taken  from \cite{Mukh2}. The solid and dashed lines are the results of the present work and the DWBA calculations of \cite{Mukh2}, respectively, for the ground ($a$), first (for $E^*$=  0.718 MeV)($b$) , second (for $E^*$= 1.740 MeV) ($c$)  and third (for $E^*$=  2.154 MeV) ($d$)  excited states of the residual nucleus ${\rm {^{10}B}}$. }
\end{figure}

\begin{figure}
\begin{center}
\includegraphics[height=12cm, width=15cm]{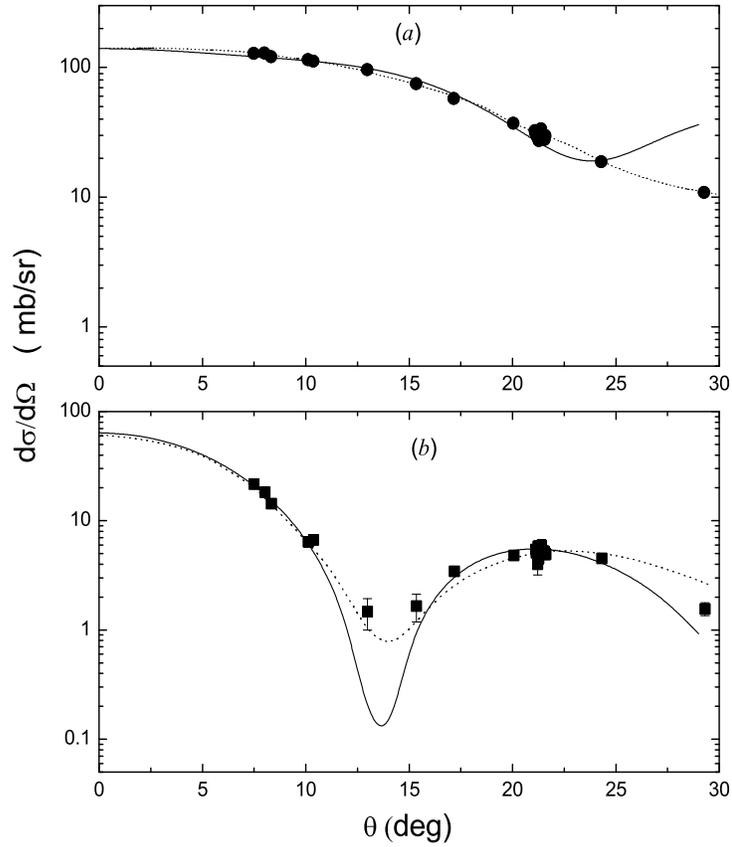}
 \end{center}
\caption{\label{fig2} The differential cross sections  for the  $ {\rm{^{16}O(^3He}}, d{\rm {)^{17}F}}$ reaction corresponding to   the ground ($a$) and first excited (0.429 MeV)($b$) states of ${\rm{^{17}F}}$   at  $E_{\rm{^3He }}$= 29.75 MeV.  The  solid and dashed curves are the results of the present work and those of Ref.  \cite{Mukh1999} derived  in  the ``post'' form of   the modified   DWBA.
 The experimental data are taken from  Refs. \cite{Mukh1999}. }
\end{figure}

\begin{figure}
\begin{center}
\includegraphics[height=18cm, width=14cm]{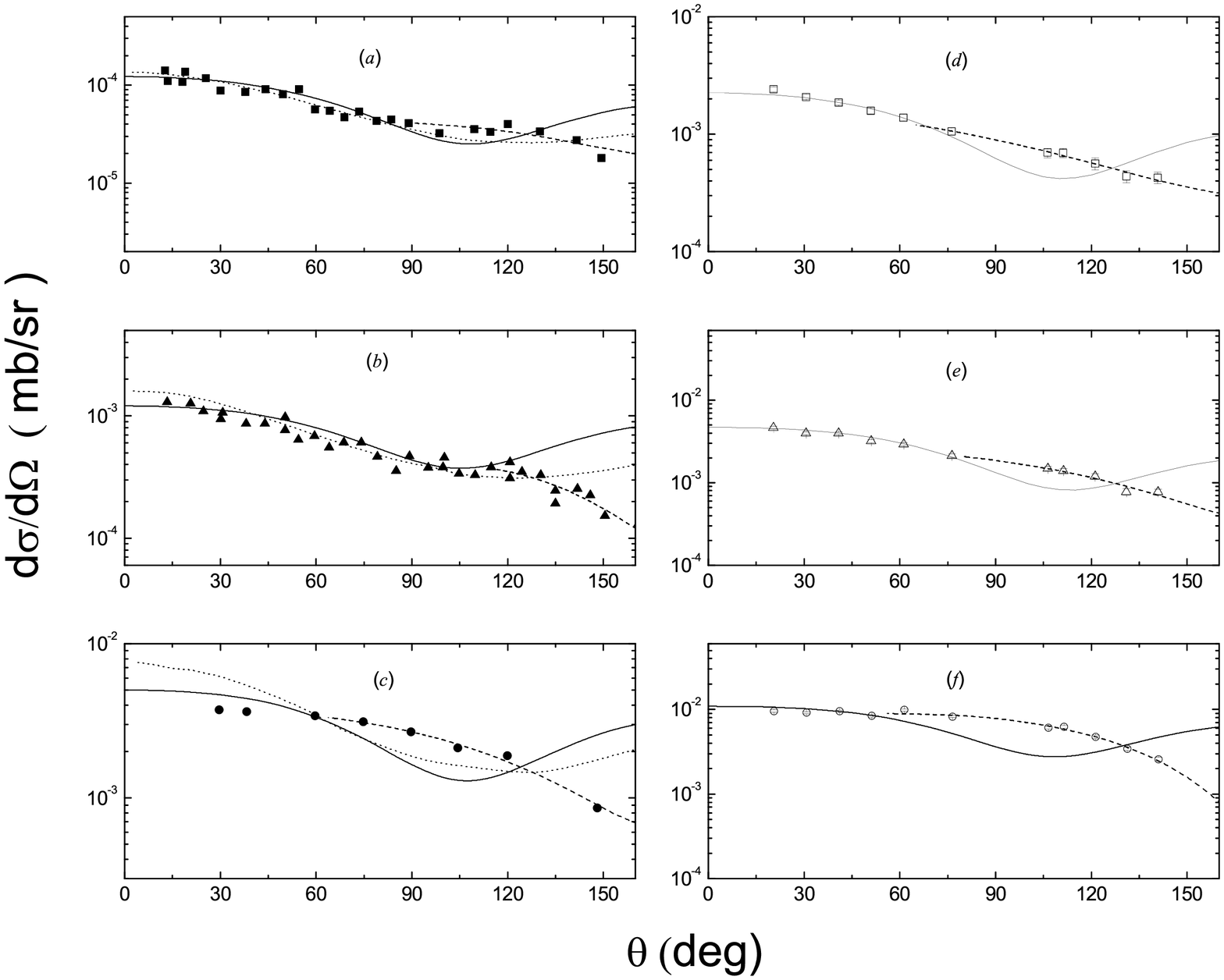}
 \end{center}
\caption{\label{fig3} The differential cross sections   for  the  ${\rm{^{19}F(}}p,\,\alpha {\rm {)^{16}O}}$ reaction  at  $E_p$=  450  ($a$),  350   ($b$) and 250 keV  ($c$) (the left side)  as well as  $E_p$= 327 ($d$),  387   ($e$) and 486 keV  ($f$) (the right side). The  solid and dotted  curves are the results of the present work, whereas the  dashed lines are the results of Ref. \cite{HAS1991} derived in the zero-range of  the ``post''-approximation of DWBA.  The dotted curves are presented our result obtained by means of the polynomial fit (see Eq. (A11) in Appendix and Table \ref{table2}).   The experimental data are taken from  Refs.  \cite{HL1978} (the EXP-1978:($a$), ($b$) and ($c$), see \cite{HAS1991} too)  and \cite{Ivano2015} (the EXP-2015:($d$), ($e$) and ($f$)).}
\end{figure}

 Figs. \ref{fig1} -- \ref{fig3} show the results of the  calculations of the  DCS's  obtained    in the present work (the solid curves), their comparison with the conventional  DWBA calculations  performed in Refs. \cite{Mukh2,Mukh1999,HAS1991}  (the dash curves) and experimental data.      The results  of the present work      correspond to the standard value of the $r_{{\rm{0}}}$ parameter, which is taken  equal to 1.25 fm and    also to   the  minimum     of $\chi^{\rm{2}}$     in  the  angular region  of  the main peak  of the angular distribution. It is seen that the angular distributions calculated in the present work   reproduce equally well the experimental data in the angular range of the main peak of the corresponding  angular distributions. The square of the ANC values and  that of  the modules of the  respective  NVC ones ($\mid G_B\mid^{{\rm{2}}}$) are  summarised in Table \ref{table1}. They   are found  by normalizing   the calculated    cross sections to  the  corresponding experimental ones at  the forward angles  and using  the ANC's $C^{\rm{2}}_{{\rm{^3He}}}$=4.20$\pm$0.32 fm$^{-1}$   ($\mid G_{{\rm{^3He}}}\mid^{\rm{2}}$= 1.32$\pm$0.10 fm) for $d\,+p\to\,{\rm{^3He}}$ compiled  in \cite{YaBl2018} and $ C_{\alpha}^{\rm{2}}$= 54.2$\pm$4.5 fm$^{-1}$\cite{Pl1973}  ($\mid G_{\alpha}\mid^{\rm{2}}$= 13.4$\pm$1.1 fm)  for  $t\,+p\to\,\alpha$. There, the   theoretical and experimental  uncertainties correspond to   variation (up to $\pm$3.0\%)
  of  the   $r_{{\rm{0}}}$ parameter with respect to     its   standard value  above and the experimental errors in $d^{\rm{exp}}\sigma/d\Omega$, respectively.
    The experimental uncertainties pointed out  in the  ANC (NVC) values for ${\rm{^{17}F}}\to\, {\rm{^{16}O }} \,+p$  and ${\rm{^{19}F}}\to\, {\rm{^{16}O }} \,+t$ correspond to  the average   squared errors, which includes both the experimental errors in $d^{\rm{exp}}\sigma/d\Omega$ and  the above-mentioned uncertainty of the ANC (NVC)  for $d\,+p\to\,{\rm{^3He}}$ and  $t\,+p\to\,\alpha$, respectively.
One notes that the value of $C^{\rm{2}}_{{\rm{^3He}}}$ ($\mid G_{{\rm{^3He}}}\mid^{\rm{2}}$) above   is in  an excellent agreement within its uncertainty with the ``indirect measured'' (``experimental'') values  of 4.28$\pm$0.50 fm$^{-1}$   (1.34$\pm$0.15 fm) \cite{Art2009} and 4.35$\pm$0.10 fm$^{-1}$   (1.36$\pm$0.03 fm) \cite{Art2008}, which were obtained from the independent indirect  methods. Therefore, they can be considered as the most reliable ones so far.

  As is seen from the first -- ninth lines of  Table \ref{table1}, the  $C_{{\rm{^{10}B_0}}}^{{\rm{2}}}$ value for ${\rm{^9Be}} \,+p\to\, {\rm{^{10}B_0}}$ obtained in the present work differs  noticeably from that of   \cite{Mukh2} derived    from the analysis of  the same reaction performed within  the framework of the  ``post'' form of the modified DWBA. This difference exceeds overall the normalization accuracy ($\Delta_{{\rm{exp}}}$= 7\% 
   \cite{Mukh2}) for the absolute values of  the  DCS's. While, such the difference  for the $C_{{\rm{^{10}B_0}}}^{{\rm{2}}}$ derived  in \cite{Mukh2} (the third and sixth lines) exceeds the $\Delta_{{\rm{exp}}}$ error      and is   about of  9\%.
 The central value of the weighed mean of the square of the  ANC  for ${\rm{^9Be}} \,+p\to\, {\rm{^{10}B_0}}$ ($C_{{\rm{^{10}B_0}}}^{{\rm{2}}}$=4.35$\pm$0.28 fm$^{-{\rm{1}}}$) recommended in the present work    is   2.5$\sigma$  lower than that of \cite{Mukh2}   presented in the ninth line of    Table \ref{table1}.
 One notes that our result
for the  weighed $C_{{\rm{^{10}B_0}}}^{{\rm{2}}}$  mean value has overall the uncertainty of  about  6\%.
 Therefore,  the value of $C_{{\rm{^{10}B_0}}}^{{\rm{2}}}$ above  is  used  by us for obtaining  the ANC's $C_{{\rm{^{10}B}}_i}^{{\rm{2}}}$ for ${\rm{^9Be}} \,+p\to\, {\rm{^{10}B}_i}$ ($i$=1--3). The results  for $C_{{\rm{^{10}B}}_i}^{{\rm{2}}}$  and their comparison with those obtained by  other authors are presented in the tenth--fifty fourth lines of Table  \ref{table1}. In Table \ref{table1},  the weighed $C_{{\rm{^{10}B}}_i}^{{\rm{2}}}$ mean values     recommended in the present work are listed
  in the  eighteenth and twenty eighth lines ($C_{{\rm{^{10}B}_1}}^{{\rm{2}}}$=1.39$\pm$0.09 and 3.74$\pm$0.32  fm$^{-{\rm{1}}}$  for $j_{{\rm{^{10}B}_1}}$=1/2 and 3/2, respectively), thirty  eighth line ($C_{{\rm{^{10}B}_2}}^{{\rm{2}}}$= 3.58$\pm$0.34  fm$^{-{\rm{1}}}$ for $j_{{\rm{^{10}B}_2}}$= 3/2) and the fourty eighth  and fifty seventh  lines ($C_{{^{10}B}_3}^{{\rm{2}}}$= 0.25$\pm$0.06 and 0.72$\pm$0.19  fm$^{-{\rm{1}}}$  for $j_{{^{10}B}_3}$=1/2 and 3/2, respectively).
There, as a comparison with the results of the present work, the results of Ref. \cite{Mukh2} are listed, which were derived using the   $C_{{\rm{^{10}B_0}}}^{{\rm{2}}}$ value presented in the ninth line. Note once more that this value is overestimated with respect to that obtained in  the present work. As is seen from Table \ref{table1},  the similar difference occurs between our results and those obtained in  \cite{Mukh2}  for the $C_{{\rm{^{10}B}}_i}^{{\rm{2}}}$ ANC's ($i$=1--3), which   is up to $\sim$19\% 
for the second exited state of   the residual ${\rm{^{10}B}}$ nucleus.
This means that the contribution of the three-body (${\rm{^9Be}},\,p$ and ${\rm{^9Be}}$) dynamics in the main pole proton transfer mechanism enhances for the excited states of ${\rm{^{10}B}}$ populating in  the entrance channel. Besides, as seen  from Table \ref{table1}, the quite notice discrepancy occurs between the results of the present work and those of  Ref. \cite{Ar96} obtained from the ${\rm{^9Be(^3He}},\,d{\rm{)^{10}B}}$ performed within the framework of the ``post''-approximation of the conventional DWBA. Therefore, it is difficult  to estimate an accuracy  of the $C_{{\rm{^{10}B}}_i}^{{\rm{2}}}$ values ($i$=1--3) derived in \cite{Mukh2,Ar96}.

  The weighed mean values      for the square of   the ANC's  for   ${\rm{^{16}O}}\,+p\to\,{\rm {^{17}F_0}}$  and  ${\rm{^{16}O}}\,+p\to\,{\rm {^{17}F_1}}$,  derived in the present work from the analysis performed   for the sets 1 and 2 of the optical potentials, are presented in the sixty seventh and eighty first lines of Table \ref{table1}, respectively.   As is seen from there, the noticeable dependence of the $C_{{\rm{^{17}F}}_i}^{{\rm{2}}}$ values ($i$=1 and 2) from the  used sets  for  the optical potentials  is observed  both for the results       of   the present work and for those of  Ref. \cite{Mukh1999}. Nevertheless, as noted above,  the results of Ref. \cite{Mukh1999} have been obtained with the underestimated $C_{{\rm{^3He}}}^{{\rm{2}}}$  value  for $d\,+p\to\,{\rm {^3He}}$ given in \cite{Mukh1995}.
  Besides, the considerable discrepancy occurs between the results of the present work and those of  Refs.\cite{Blokh2018} and \cite{Ar96} obtained within the continuation method for the experimental $p{{\rm{^{16}O}}}$-scattering function and the ${\rm{^{16}O(^3He}},\,d{\rm{)^{17}F}}$ DCS analysis performed within the finite range of the  ``post''-approximation of the modified DWBA, respectively. By using this case, one notes that there are misprints in the first line of Table I of Ref. \cite{Blokh2018}. There, the figures of 75.5$\pm$15 and 1.1$\pm$0.33 fm$^{{\rm{-1/2}}}$ corresponding respectively for  the ANC's for ${\rm{^{16}O}}\,+p\to\,{\rm {^{17}F_1}}$  and  ${\rm{^{16}O}}\,+p\to\,{\rm {^{17}F_0}}$  must be replaced by 75.5$\pm$1.5 and 1.04$\pm$0.05 fm${^{\rm{-1/2}}}$, respectively.
      Nevertheless, as is seen from Table \ref{table1}, our results for the $C_{{\rm{^{17}F}}_i}^{{\rm{2}}}$ above    are in good agreement  within about 1$\sigma$ with those of Refs. \cite{Art2009} and \cite{YarBaye2011}, which were derived by the quite other ``indirect'' methods.

  The $C_{{\rm{^{19}F}}}^{{\rm{2}}}$   values for
    ${\rm{^{16}O}} \,+\,t\to \,{\rm{^{19}F}}$ obtained in the present work at the different proton projectile  energies and their weighted means  are presented  the eighty eighth -- ninety eighth    lines of Table   \ref{table1}. As it is seen from  there, the weighted $C_{{\rm{^{19}F}}}^{{\rm{2}}}$ mean values    found   separately from the analysis of  the    experimental data taken from Ref. \cite{HL1978} (EXP-1978) and from Ref. \cite{Ivano2015} (EXP-2015), which are listed in the ninety third and ninety eighth lines of Table \ref{table1}, respectively, differ from each other on the average  by  a factor of about  2.2.  This is the  result of the discrepancy between  the absolute  values of the experimental DCS's of  the EXP-1978 and the EXP-2015 measured  independently at fairly close energies.  To find out  the main reason of   this   discrepancy, we recommend decisive measurement of the experimental DCS's of the   ${\rm{^{19}F(}}p,\,\alpha {\rm {)^{16}O}}$  reaction  in the sub-barrier projectile energy region being  closer to that of Refs. \cite{Ivano2015,HL1978}.  Nevertheless, one notes that the $C_{{\rm{^{19}F}}}^{{\rm{2}}}$ value   obtained separately  from the independent experimental data of the  EXP-1978 and EXP-2015  at  the different projectile energies are stable, although the absolute  values of the corresponding experimental DCS's of  the  EXP-1978 and EXP-2015  depend strongly   on  the proton projectile energy (see Fig. \ref{fig3}). This result   confirms    the assumption made in Ref. \cite{YTur2019} about  possibility of the applicability of the asymptotic theory developed in  \cite{YTur2019} also  for the sub-barrier  peripheral charged-particle transfer reactions  as a tool of obtaining the ANC.
    To best of our knowledge,  the ANC value for   ${\rm{^{16}O}} \,+\,t\to\, {\rm{^{19}F}}$ presented in   Table \ref{table1} are obtained for the first time.

    As it is seen from  the analysis performed   above,     the asymptotic theory proposed in Ref.
    \cite {YTur2019} provides better  accuracy for the ANC values   for $ {\rm{^9Be}} \,+p\to\,{\rm{^{10}B}}$   and ${\rm{^{16}O}}\, +\,p\to {\rm {^{17}F}}$  than those obtained  in Refs. \cite{Mukh2,Mukh1999,Ar96} for their nuclear-astrophysical application. Besides, the  ANC values  for    ${\rm{^{16}O}} \,+\,t\to\, {\rm{^{19}F}}$ derived above can give a valuable information about the astrophysical $S$ factors (or the cross sections) for the ${\rm{^{19}F(}}p,\,\alpha{\rm{)^{16}O}}$ reaction in the astrophysically relevant energy region where  the direct transfer  mechanism is dominant. This issue is considered below.
 \bigskip

   {\bf B.  Astrophysical $S$ factors at stellar energies}

\bigskip

 Here the weighted mean values of  the ANC's obtained by us for $ {\rm{^9Be}} \,+p\to\,{\rm{^{10}B}}_i$ ($i$=0--3) and  ${\rm{^{16}O}}\, +\,p\to\, {\rm {^{17}F_i}}$ ($i$=0 and 1)  are used to calculate the   astrophysical $S$ factors for the   radiative capture  ${\rm{^9Be(}}p,\,\gamma{\rm{)^{10}B}}$ and  ${\rm{^{16}O(}}p,\,\gamma{\rm{)^{17}F}}$     reactions at stellar energies.  The calculations are performed  within the framework of    the modified $R$-matrix  method  (see Ref. \cite{Ar2012} for example) for the radiative capture  ${\rm{^9Be(}}p,\,\gamma{\rm{)^{10}B}}$ reaction, where  the direct component of the total amplitude   is determined by solely the ANC values, and of the modified two-body potential method (MTBPM) \cite{Igam07} for the  direct radiative capture ${\rm{^{16}O(}}p,\,\gamma{\rm{)^{17}F}}$ reaction, where the direct astrophysical $S$ factor is parameterized in the term of the square of the ANC's above. For easer  reading of the paper, the basic formulas of these methods  are given in Appendix.  Besides, the ANC's derived for ${\rm{^{16}O}} \,+\,t\to\, {\rm{^{19}F}}$(g.s.) are used for getting information about the astrophysical $S$ factors of the nuclear-astrophysical ${\rm{^{19}F(}}p,\,\alpha {\rm {)^{16}O}}$ reaction at six the  proton energies mentioned above by means of the way   presented in Appendix.

The analysis  of the experimental astrophysical $S$ factors  for the  ${\rm{^9Be(}}p,\,\gamma{\rm{)^{10}B}}$  reaction   was performed  by  taking into account the contributions from  captures to fours the resonances and to the direct capture as well as their interference contributions. One notes that  three from these resonances are broad ones at $E^{{\rm{(R)}}}_{{\rm{1}}}$=287 keV with $J^{\pi}$=1$^-$,  $E^{{\rm{(R)}}}_{{\rm{2}}}$=892 keV with $J^{\pi}$=2$^+$ and $E^{{\rm{(R)}}}_{{\rm{4}}}$=1161 keV with $J^{\pi}$=2$^-$, and one from them  is narrow resonance at  $E^{{\rm{(R)}}}_{{\rm{3}}}$=975 keV with $J^{\pi}$=0$^+$ \cite{Ajzen1988}.     As in Refs. \cite{Satt1999,Zah1995}, we consider   the  following  transitions from the  resonant states above to the ground and   three  first excited bound states of the residual ${\rm{^{10}B}}$ nucleus: the 0$^+$ third resonance $\to$ the first (1$^+$) and third (1$^+$) bound excited  states of ${\rm{^{10}B}}$; the first (1$^-$), second (2$^+$) and fourth (2$^-$) resonances $\to$    all the considered bound (ground--excited)  states of ${\rm{^{10}B}}$. The calculation is performed using Eqs. (A1)--(A8) given in Appendix. As is seen from Eq. (A8),   the power of the direct amplitude of the total  amplitude given by Eq. (A4) is determined only the  $C_{{\rm{^{10}B}}_i}$ ANC's found above, which will be   used in   calculations below. Besides, as is seen from Eqs. (A3) and (A4) of Appendix,    the direct and resonance terms with  the same channel spin ($I$=1 or 2)  interfere only with each other. The calculations show that the resonant  and   direct amplitudes  are  formed  predominantly by  the $p{\rm{^9Be}}$-scattering $s$ wave capture and the  $E$1 capture, respectively.  The direct $M$1 and $E$2 contributions into the direct amplitude for all the transitions in the exit channel  are negligibly small with respect to the dominant $E$1 contribution. Hence, they can be ignored.

\begin{figure}
\begin{center}
\includegraphics[height=12cm, width=14cm]{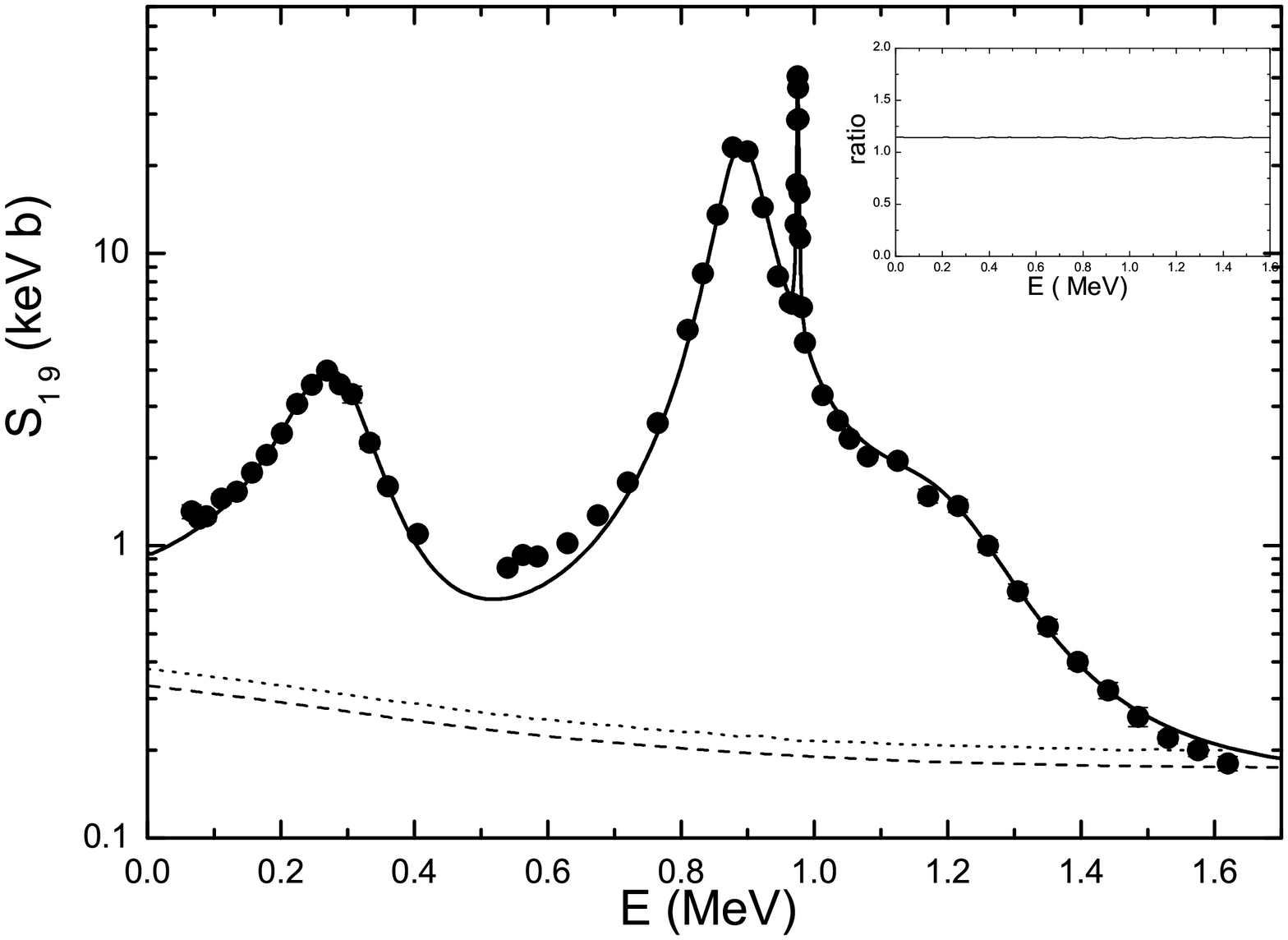}
 \end{center}
 \vspace{-2mm}
\caption{\label{fig4} {\small The total  astrophysical $S$ factor for the ${\rm{^9Be(}}p,\,\gamma{\rm{)^{10}B}}$ reaction. The point are the experimental data from \cite{{Zah1995}}. The solid  and dashed lines  are the calculated results of the present work for the total and direct radiative capture, respectively. The curve  in the insert is the ratio of the direct component of $S_{{\rm{1\,9}}}$($E$) of Ref. \cite{Satt1999} to that of the present work.   }}
     \end{figure}

Fig. \ref{fig4} shows the results of the calculations  for    the  total ($S_{{\rm{1\,9}}}$($E$) plotted by the solid curve) and direct ($S_{{\rm{1\,9}}}^{{\rm{DC}}}$($E$) plotted by the dashed curve)  astrophysical $S$ factors, which are in  an excellent agreement with the experimental data \cite{Zah1995}. This is connected apparently with the correct taking into account of the energy dependence of the $\gamma$-width, which contains both the interior contribution and the channel one defined  by Eq. (A6b) of Appendix.    As seen from Fig. \ref{fig4}, the noticeable difference  occurs between the direct component  of $S_{{\rm{1\,9}}}$($E$) derived in Ref. \cite{Satt1999} (the dashed-dotted line) and that obtained in the present work (the dashed line). See also the inset there. Their ratio changes from 1.10 to 1.14 with an increase of the energy $E$. This is due to  the overestimated values of the ANC's compared to those of the present work (see Table \ref{table1}).
 The fitted parameters of all the resonance levels are given in Table \ref{table2}.
The value of the channel radius $r_{c}$  ($r_{c}$= 3.1 fm) is chosen to provide the minimum of $\chi^{{\rm 2}}$  ($\chi^{{\rm 2}}$=2.5) in fitting data, which is noticeably less than that ($\chi^{{\rm 2}}$=7.8) obtained in Ref.\cite{Satt1999}. In the calculations, the  $\gamma$-widths of  the resonances were considered as  adjustable parameters. The protonic- and $\alpha$-channel widths for the ground and first three excited states of ${\rm{^{10}B}}$ are taken from  Ref. \cite{Ajzen1988}. As is seen from Table \ref{table2}, the absolute values of the
$\gamma$-widths for the first, second and fourth resonances found  in the present work by using Eq. (A7), are in  good agreement with the results of \cite{Ajzen1988}, except for  the $\gamma$-width for the third (0$^+$) resonance. The $\gamma$-width value for the 0$^+$ resonance  is found to be   $\Gamma^{\gamma}$=6.5 eV, which   differs about 31\% 
from that ($\Gamma^{\gamma}=$8.5 eV)  recommended in \cite{Ajzen1988}. Note that, in \cite{Satt1999}, all the $\gamma$-width values   were fixed   and taken from \cite{Ajzen1988}.   As is seen from Fig. \ref{fig4}, the calculated total  astrophysical $S$ factor reproduces   fairly good  the experimental data.  In particular, the $S_{{\rm{1\,9}}}$(0)=0.946$\pm$0.194 keV$\cdot$b and $S_{{\rm{1\,9}}}^{{\rm{DC}}}$(0)= 0.331$\pm$0.021 keV$\cdot$b  as well as   $S_{{\rm{1\,9}}}$(25 keV)= 0.970$\pm$0.200 keV$\cdot$b and
$S_{{\rm{1\,9}}}^{{\rm{DC}}}$(25 KeV)= 0.327$\pm$0.021 keV$\cdot$b are obtained.
The value  $S_{{\rm{1\,9}}}^{{\rm{DC}}}$(0) derived in the present work is 2.5$\sigma$   lower than that of  $S_{{\rm{1\,9}}}^{{\rm{DC}}}$(0)= 0.38$\pm$0.02 keV$\cdot$b \cite{Satt1999,Wulf1998}.  This difference is associated  with the model assumptions used in Refs. \cite{Satt1999,Wulf1998}.  Nevertheless, our result derived  for the total $S_{{\rm{1\,9}}}$(0) is in good agreement with that of
   $S$(0)=0.96$\pm$0.02  keV$\cdot$b, $S$(0)=0.96$\pm$0.06  keV$\cdot$b  and $S_{{\rm{1\,9}}}$(0)=1.0$\pm$0.1 keV$\cdot$b derived in Refs. \cite{Wulf1998}, \cite{Satt1999} and \cite{Zah1995}, respectively.  
    
    The calculations of the  astrophysical $S$ factors    for the direct   radiative capture     ${\rm{^{16}O(}}p,\,\gamma{\rm{)^{17}F}}$  reaction  are done by using Eq. (A9) of Appendix. In Eq. (A9),  the weighted mean  values  of the $C_{{\rm{^{17}F}}_i}^{{\rm{2}}}$ ANC's ($i$=1 and 2)  for ${\rm{^{16}O}}\,+p\to\,{\rm {^{17}F_1}}$  and  ${\rm{^{16}O}}\,+p\to\,{\rm {^{17}F_0}}$ derived in the present work  are used. Whereas, the ${\cal{R}}_{l_{{\rm{^{17}F}}_j}}{\rm{(}}E;\,b_{{\rm{^{17}F}}_j}{\rm{)}}$ function  is calculated similarly to that  as is done in Ref. \cite{Art2009}. Nevertheless,  we note only the following. The direct amplitude of the reaction above  is formed predominantly through   the $E$1,  $M$1  and $E$2 capture. For the transition to the ground (${\rm {^{17}F_0}}$) state,     the  $p$, $f$($d$)   waves and   the $s$ and $d$ ones correspond respectively to     the $E$1($M$1) and $E$2 transitions.  For  the transition to the first  (${\rm {^{17}F_1}}$) excited  state,  the $p$($s$) wave and the $d$ wave correspond  respectively to   the same transitions above. Besides, as shown in \cite{Art2009}, the uncertainty of the calculated   ${\cal{R}}_{l_{{\rm{^{17}F}}_j}}{\rm{(}}E;\,b_{{\rm{^{17}F}}_j}{\rm{)}}$ function is about   $\pm$4\%. It arises under the  variation of the
  free $b_{{\rm{^{17}F}}_j}{\rm{(}}r_{{\rm{0}}},\, a{\rm{)}}$ parameter
 relatively its value corresponding to  the standard  ($r_{{\rm{0}}}$=1.25 fm and $a$=0.65 fm) values of the geometric  parameters of the adopted Woods-Saxon potential. Note that  this potential  is used  for calculating both the bound ${\rm{^{17}F}}_j$  state wave function and the continuum $p{\rm{^{16}O}}$-scattering one, which   enter  the radial integral of the matrix element \cite{Igam07}.

\begin{figure}
\begin{center}
\includegraphics[height=11cm, width=9cm]{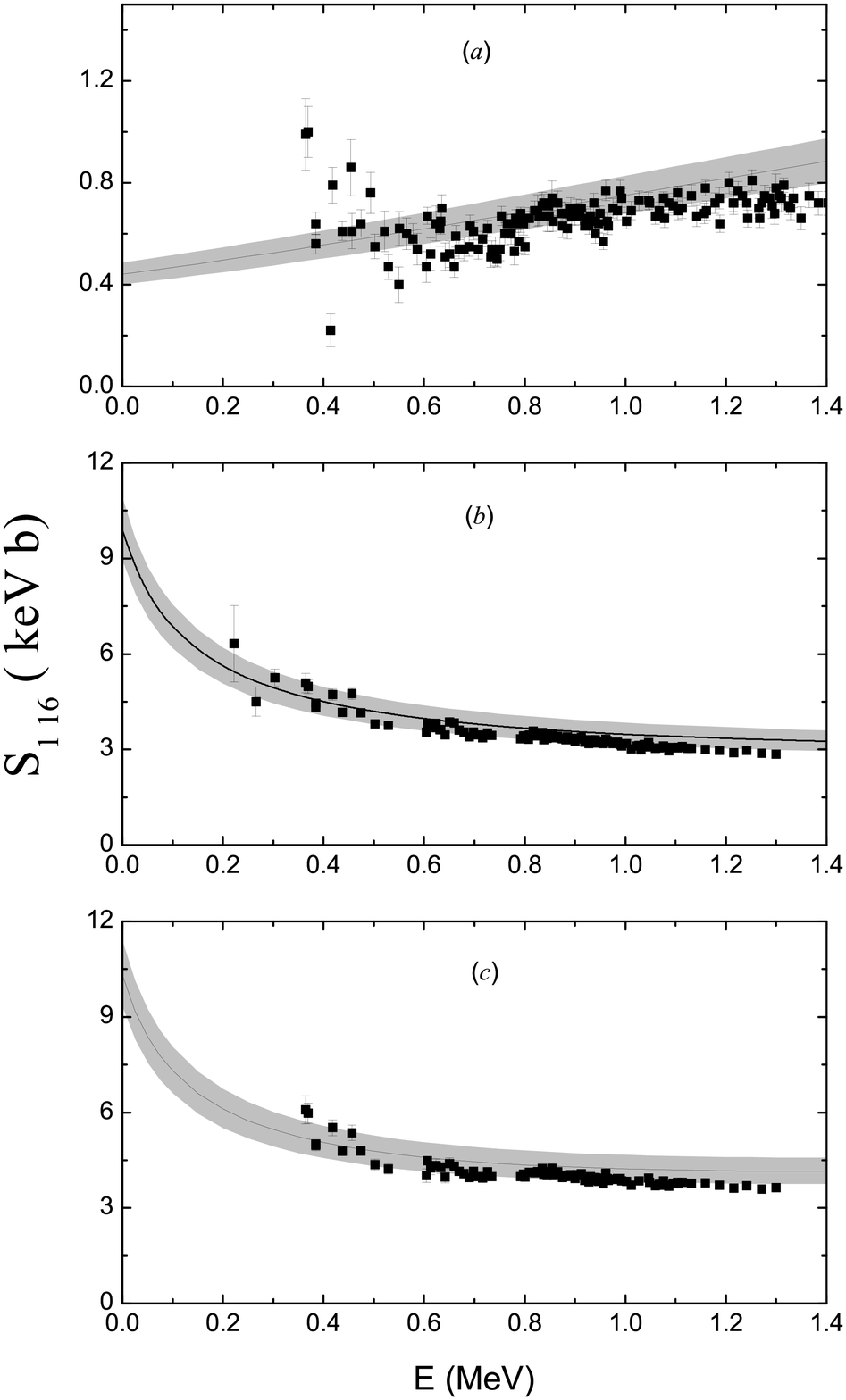}
\end{center}
\vspace{-2mm} \caption{\label{fig5}{\small The
astrophysical $S$ factors for the direct radiative capture ${\rm {^{16}O(}}p,\gamma{\rm {)^{17}F}}$ reaction. The curves of  ($a$) and ($b$) correspond to  the ground and  first excited (0.495 MeV) states  of the residual ${\rm{^{17}F}}$ nucleus, respectively, whereas  that of  ($c$) corresponds to  their sum
${\rm{^{17}F}}$ (g.s. + 0.495 MeV).
  The solid  and  the band  are  the results of
the present work, whereas   the  dashed  line is the result of   Ref. \cite{Art2009}. The experimental data are  from \cite{Morl1997}.}}
\end{figure}

     The results of  comparison  between the   astrophysical $S$ factors ($S_{{\rm{1\,16}}}$($E$)) calculated in the  present work  and the experimental data  \cite{Morl1997} are displayed in Fig. \ref{fig5}.
     There, the  solid curves in  ($a$) and ($b$)   present the results    for    the ground and  first excited  states   of  the residual ${\rm{^{17}F}}$ nucleus, respectively, whereas  the solid curve in ($c$) corresponds to  their sum
   ${\rm{^{17}F}}$(g.s. + 0.429 MeV).
    In Fig. \ref{fig5}, the width of the bands are the  uncertainties, which are the average squared errors  of the uncertainties of the ANC's given in Table \ref{table1} and that of the ${\cal{R}}_{l_{{\rm{^{17}F}}_j}}$ function mentioned above, and the  dashed curves   are the results   of Ref. \cite{Art2009}.   As is seen from figure, the weighted $C_{{\rm{^{17}F}}_i}^{{\rm{2}}}$ mean  values  derived  in the present work firstly, reproduce well the  experimental data and, secondly,   allow extrapolation  of the astrophysical $S$ factors ($S_{{\rm{1\,16}}}$($E$)) at stellar energies.
 In a particular,  $S_{{\rm{1\,16}}}^{{\rm{g.s.}}}$($E$)= 0.44$\pm$0.04  and 0.45$\pm$0.05 keV$\cdot$b as well as
   $S_{{\rm{1\,16}}}^{{\rm{exc.}}}$($E$)= 9.89$\pm$1.01  and 9.20$\pm$0.94 keV$\cdot$b are obtained for $E$= 0 and 25 keV, respectively. And, the
 total astrophysical $S$ factors  $S_{{\rm{1\,16}}}$($E$) are found to be  10.34$\pm$1.06  and 9.65$\pm$0.98 keV$\cdot$b   for $E$= 0 and 25 keV, respectively. One notes that our result   for $E$= 0  agrees     with that of  $S_{{\rm{1\,16}}}$(0)=9.45$\pm$0.4 keV$\cdot$b   \cite{Art2009}  and  with the results of  10.2 and 11.0 keV$\cdot$b  \cite{BD1998} obtained  within the framework of the microscopic model for the effective  V2 and MN  potentials of the NN potential, respectively.

\begin{figure}
\begin{center}
\includegraphics[height=11cm, width=12cm]{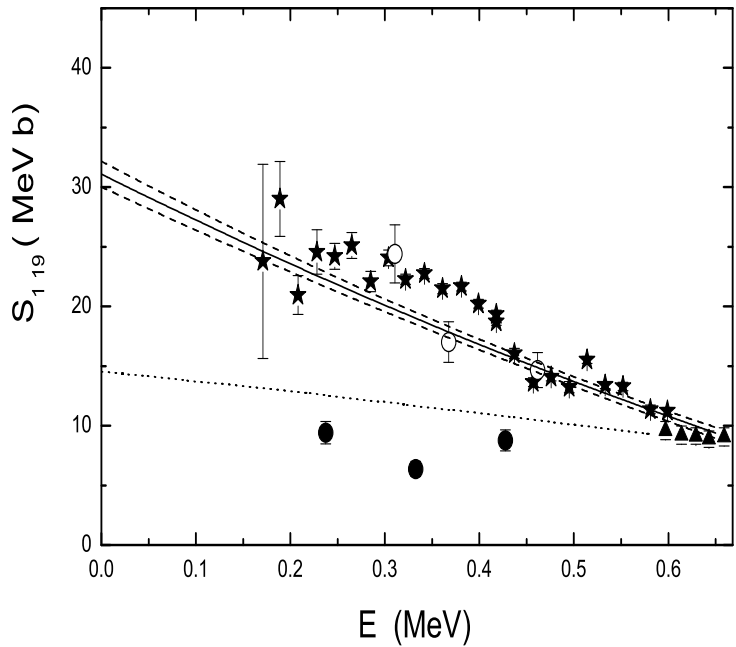}
\end{center}
\vspace{-2mm} \caption{\label{fig6}{\small The
astrophysical $S$ factors for the direct radiative capture ${\rm {^{19}F(}}p,\alpha{\rm {)^{16}O}}$ reaction: open and full cycle points are our result for the direct $S_{{\rm{1\,19}}}$($E$) astrophysical $S$ factor obtained from the analysis of the  EXP-1978 \cite{HL1978,HAS1991} and EXP-2015 \cite{Ivano2015} data;
 The experimental star points are taken from Refs \cite{Ivano2015,Con2015} (stars points) and  \cite{Isoya1958} (full triangle points). The solid curve   corresponds to  the  polynomial fitting.
 The lower dotted curve is the results of Ref. \cite{An99} obtained from the  linear extrapolated formula used   for the experimental $S_{{\rm{1\,19}}}{\rm{(}}E{\rm{)}}$ data   of \cite{Isoya1958}.
  }}
\end{figure}

In Fig. \ref{fig6}, the results for the astrophysical $S$ factors ($S_{{\rm{1\,19}}}$($E$)) for  the nuclear-astrophysical ${\rm{^{19}F(}}p,\,\alpha {\rm {)^{16}O}}$ reaction, obtained  in the present work for corresponding  six    proton energies mention above, are displayed  by open and full cycle points. They were obtained from the expressions (A1), (A10) and A(11) of Appendix with the fitted coefficients $a_n{\rm{(}}E{\rm{)}}$ (0$\le n\le$2) from Table \ref{table2}   and the corresponding   ANC $C_{{\rm{^{19}F}}}^{{\rm{2}}}$ values for ${\rm{^{16}O}} \,+\,t\to\, {\rm{^{19}F}}$ from Table \ref{table1}. There, the uncertainty  for each the energy $E$  corresponds respectively to that of the  $C_{{\rm{^{19}F}}}^{{\rm{2}}}$ ANC.
Open  and full cycle points in Fig. \ref{fig6} correspond  respectively to  the  $C_{{\rm{^{19}F}}}^{{\rm{2}}}$  values obtained in the present work   from the analysis of  the  EXP-1978 data \cite{HL1978} for $E$=237.5; 332.5 and  427.5 keV and the  EXP-2015 data \cite{Ivano2015} for $E$=310.7; 367.7 and  461.7 keV.  The  experimental data plotted in  Fig. \ref{fig6} by star points  are taken  from Refs. \cite{Ivano2015,Con2015}.  As is seen from this figure, the  open cycle data    are in a reasonable agreement with those of Refs. \cite{Ivano2015,Con2015}, whereas the full cycle data differ noticeably from them. This discrepancy  is   mainly the result of the fact that the corresponding $C_{{\rm{^{19}F}}}^{{\rm{2}}}$ values used in Eqs. (A10a) and (A10b) are underestimated by a factor of about 2  compared to those obtained from the analysis of the EXP-2015 \cite{Ivano2015}.  Therefore, the present complex analysis performed on the basis of the asymptotic theory developed in \cite{YTur2019} gives strong evidence that   ${\rm{^{19}F(}}p,\,\alpha {\rm {)^{16}O}}$  astrophysical $S$ factors   at the   considered thermonuclear energies, which   belong to  the  energy    region      considered in Refs. \cite{Ivano2015,Con2015},  is  dominant by the one-step    triton transfer pole mechanism in which  the  three-body Coulomb dynamics in the main transfer mechanism is   taken into account   correcty.

 On the other hand, the data for $S_{{\rm{1\,19}}}$($E$) plotted in Fig. \ref{fig6} by open cycles can be parametrized   in the analytical form
\begin{equation}
 S_{{\rm{1\,19}}}{\rm{(}}E{\rm{)}}={\rm{(31.09}}\pm{\rm{1.08)}}-{\rm{(39.44}}\pm{\rm{2.53)}}E+
 {\rm{(9.34}}\pm{\rm{2.41)}}E^{{\rm{2}}},
\label{Sfact}
\end{equation}
in which  $ S_{{\rm{1\,19}}}{\rm{(}}E{\rm{)}}$ in  MeV b and $E$  in MeV. Herein,
 the uncertainties in the coefficients of the polynomial expansion correspond to those of the fitted data within their errors. In Fig. \ref{fig6}, the solid curve corresponds to  the calculated polynomial approximation of  Eq.  (\ref{Sfact}) for the central values of its coefficients. While,   the upper and lower dashed curves are the results of the calculation of  Eq.  (\ref{Sfact})  with the  coefficients corresponding  respectively to  their upper and lower values. As seen from this figure, the polynomial approximation  (\ref{Sfact})   reproduces  reasonably  also the absolute values of  the data of  \cite{Ivano2015}. Therefore, Eq. (\ref{Sfact}) can be used for calculation of  $S_{{\rm{1\,19}}}{\rm{(}}E{\rm{)}}$ in the nonresonant energy range, where the direct mechanism is dominant, including  at $E\le$ 50 keV. In particular, $S_{{\rm{1\,19}}}{\rm{(}}E{\rm{)}}$=31.09$\pm$1.08,   30.11$\pm$1.08 and 29.14$\pm$1.09 MeV b for $E$= 0, 25 and 50 keV, respectively. They are significantly larger than both $S_{{\rm{1\,19}}}{\rm{(}}E{\rm{)}}$=8.76,  8.67 and 8.61 MeV b for $E$= 0, 25 and 50 keV, respectively, derived in \cite{HAS1991} from the DWBA analysis of the  EXP-1978 data  \cite{HL1978} and   the corresponding values reported in  NACRE  \cite{An99} plotted in Fig. \ref{fig6} by the dotted curve. The latter   were derived from the nonresonant   linear extrapolated formula used   for the experimental $S_{{\rm{1\,19}}}{\rm{(}}E{\rm{)}}$ data \cite{Isoya1958} from   the fairly narrow energy range of   $E\sim$ 600 keV (see the full treangle points in Fig. \ref{fig6}). One notes that the $S_{{\rm{1\,19}}}{\rm{(}}E{\rm{)}}$ data of Ref. \cite{Isoya1958} are noticeably underestimated as a comparison with those of \cite{Ivano2015}.

 \bigskip

  {\bf VII. THERMONUCLEAR   ${\rm{^9Be}(}p,\,\gamma{\rm{)^{10}B}}$ AND ${\rm{^{16}O}(}p,\,\gamma{\rm{)^{17}F}}$   REACTIONS  RATES}
  \bigskip

   The new values obtained for the total astrophysical $S$ factors for the  ${\rm{^9Be}(}p,\,\gamma{\rm{)^{10}B}}$ and ${\rm{^{16}O}(}p,\,\gamma{\rm{)^{17}F}}$  reactions were used to calculate the rates   of these reactions   as a function of stellar temperature  within the range
of 10${\rm{^{-3}}}\leq T_{\rm{9}}\leq$10, where $T_{\rm{9}}$ is a temperature  in unit of
10$^{\rm{9}}$ K.   The Maxwellian-averaged reaction rates $N_A\langle\sigma_{ij}v_{ij}\rangle$ are given by \cite{Fowler1984,Rolf88}
\begin{equation}
N_A\langle\sigma_{ij}v_{ij}\rangle=N_A\left( \frac{8}{\pi \mu_{ij}}\right )^2(k_BT)^{-3/2}
\int_0 ^{\infty} S_{ij}{\rm{(}}E{\rm{)}}exp[-E/k_BT-2\pi\eta_{ij}]dE
\label{rates}
\end{equation}
as a function of the temperature $T$.
 Herein $N_A$ is the Avogadro number; $k_B$ is the Boltzmann constant; $\mu_{ij}$ and $\eta_{ij}$
  =$Z_iZ_je^2/\hbar v_{ij}$  are the reduced mass and the Coulomb paramerter for  the colliding
($i$ and $j$) particles, respectively,     and    $v_{ij}=\sqrt{{\rm{2}}E/\mu_{ij}}$,
       where  $Z_ke$ is the charge of the particle $k$.

        To calculate  the $N_A\langle\sigma_{{\rm{{1\,16}}}}v_{{\rm{1\,16}}}\rangle$ rate for the  ${\rm{^{16}O}(}p,\,\gamma{\rm{)^{17}F}}$
        reaction, the contributions of the resonant ($E^*$=3.104 MeV with $J^{\pi}$=$ \frac{1}{2}^-$ and $E^*$=3.851 MeV with $J^{\pi}$=$ \frac{5}{2}^-$) states of ${\rm{^{17}F}}$ to the total astrophysical $S$ factor $S_{{\rm{1\,16}}}$($E$) have been taken into account within the modified $R$-matrix method,  similar to that as it is done for the ${\rm{^9Be(}}p,\,\gamma{{\rm)^{10}B}}$ reaction above.  At this, we considered the transitions from the  first and second  resonant states mentioned above to the first excited and ground bound states of the residual $ {\rm{^{17}F}}$ nucleus, respectively \cite{Morl1997}.   Both the transitions correspond to the dominant
       $E$1 and $M$1 onces.  The  values of the $\gamma$-widths  corresponding to  the  first and second resonances  as well as of the proton and total widths are taken from \cite{RBF1996}.  The direct component of $S_{{\rm{1\,16}}}$($E$) is determined by the corresponding ANC values  for  $  {\rm{^{16}O}}+p\to{\rm{^{17}F}}$ derived in the present work.  Besides, we calculate the reaction rate without taking into account the resonance contribution to the amplitude reaction. Both the method gave practically  the same results for the $N_A\langle\sigma_{{\rm{{1\,16}}}}v_{{\rm{1\,16}}}\rangle$ rate.

\begin{figure}
\begin{center}
\includegraphics[height=13cm, width=11cm]{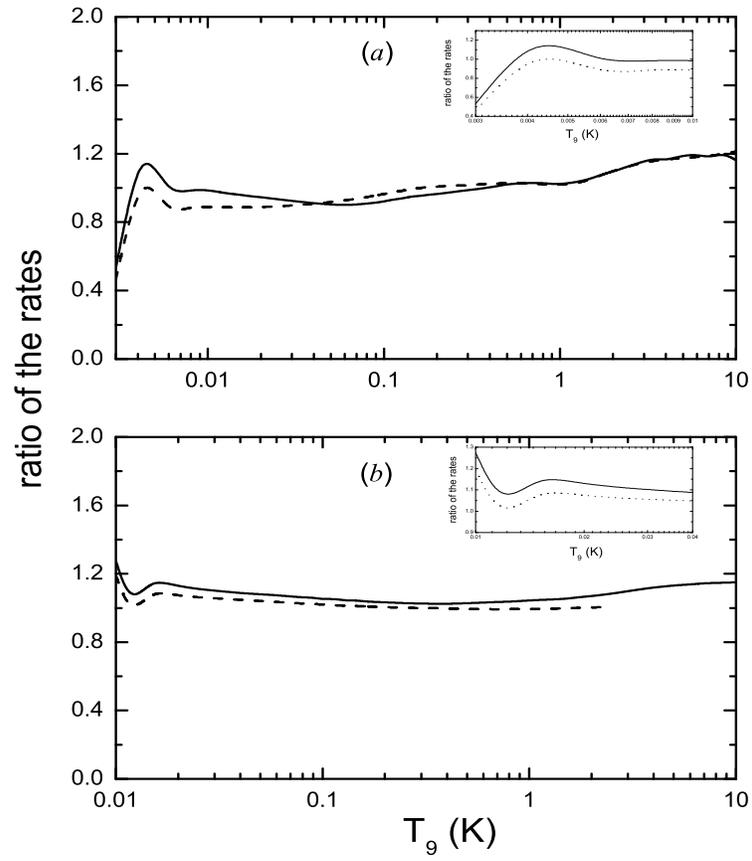}
\end{center}
\vspace{-2mm} \caption{\label{fig7}{\small The ratios  of the
rates   for the  radiative capture ${\rm {^9Be(}}p,\gamma{\rm {)^{10}B}}$ ($a$) and  ${\rm {^{16}O(}}p,\gamma{\rm {)^{17}F}}$  ($b$) reactions. Description  is given in the text.  }}
\end{figure}

The resulting numerical values of the ${\rm{^9Be}(}p,\,\gamma{\rm{)^{10}B}}$ and ${\rm{^{16}O}(}p,\,\gamma{\rm{)^{17}F}}$    reaction  rates   in the temperature range 0.01$\le T_{\rm{9}}\le$10 K are presented in Table \ref{table3}.
              Fig. \ref{fig7}($a$)   shows the ratios of the rates for the  ${\rm{^9Be}(}p,\,\gamma{\rm{)^{10}B}}$ reaction  calculated in the present work to those of Ref. \cite{An99} (the solid curve) and \cite{NACREII} (the dashed line). Whereas,   those for the  ${\rm{^{16}O}(}p,\,\gamma{\rm{)^{17}F}}$    reaction  calculated in the present work to those of Ref. \cite{An99} (the solid curve) and \cite{IIiadis2008} (the dashed line) are displayed in Fig. 7($b$). The calculations show that, at temperatures  $T_{\rm{9}}\gtrsim$ 0.004 K,       the difference between  the  calculated rates  recommended in  the present work and those   recommended in \cite{An99} and \cite{NACREII}  is noticeable, whereas, it inceases with  decreasing  the temperature (see the inserts in Fig. \ref{fig7}).
   One of the possible reasons of this discrepancy can apparently be associated with the model assumption used in \cite{An99,NACREII,IIiadis2008}, in particular,  with the choice of  the values of the spectroscopic factors for the ${\rm{^{10}B}}$ in the (${\rm{^9B}}+p$) configuration  and the ${\rm{^{17}F}}$  in the   (${\rm{^{16}O}}+p$) one, which really are strongly model dependent \cite{Art2009,Mukh2} as noted above.

       \bigskip

  {\bf VIII.  CONCLUSION }
  \bigskip

   \hspace{0.6cm}  Within the  asymptotic theory    proposed in \cite{YTur2019} for the  peripheral  sub- and above-barrier charged-particle transfer  $A$($x$, $y$)$B$ reaction, where  $x$=($y$ + $a$),  $B$=($A$ + $a$) and  $a$ is the transferred particle, the analysis   is performed  for    the  experimental angular distributions of the differential cross sections of the   specific peripheral proton and triton transfer  reactions at  above- and sub-barrier projectile energies, respectively.  It is demonstrated   that     the asymptotic theory       gives   an adequate description  both of  the  angular distributions   in  the angular region of the    main peaks of the angular distributions  and   of the absolute   values of the specific  ANC's (NVC's). New values  and   their uncertainties are  obtained for the square of the ANC's
for $ {\rm{^9Be}} \,+p\to\,{\rm{^{10}B}}$,  ${\rm{^{16}O}}\, +\,p\to {\rm {^{17}F}}$  and ${\rm{^{16}O}} \,+\,t\to\, {\rm{^{19}F}}$.
 The accuracy of   the ``post''-approximation  and the ``post'' form of the  conventional DWBA is estimated for the ANC values for $ {\rm{^9Be}} \,+p\to\,{\rm{^{10}B}}$
  and  ${\rm{^{16}O}}\, +\,p\to {\rm {^{17}F}}$ obtained  by other authors in Refs. \cite{Ar96} and \cite{Mukh2}, respectively.
  The ANC values obtained in the present work  were then applied  for calculations of   the     astrophysical  $S$ factors   for the   radiative capture   ${\rm{^9Be(}}p,\,\gamma{\rm{)^{10}B}} $ and ${\rm{^{16}O(}}p,\,\gamma{\rm{)^{17}F}}$ reactions at stellar energies as well as  of  the nuclear-astophsysical
 ${\rm{^{19}F(}}p,\,\alpha{\rm{)^{16}O}}$ reaction at  sub-barrier energies.  New  values and their uncertainties   are obtained for the astrophysical $S$ factors at thermonuclear energies. It is shown that the present analysis gives strong evidence that ${\rm{^{19}F(}}p,\,\alpha {\rm {)^{16}O}}$   astrophysical $S$ factors (or  respective cross sections) at   energies within  the range of  238$\lesssim E\lesssim$462 keV  is  dominant by the one-step    triton transfer pole mechanism in which  the  three-body Coulomb dynamics in the main transfer mechanism is   taken into account  in a correct manner. And, new values of the rates of
 ${\rm{^9Be(}}p,\,\gamma{\rm{)^{10}B}} $ and ${\rm{^{16}O(}}p,\,\gamma{\rm{)^{17}F}}$ reactions were obtained at stellar temperature  within the range
of 10${\rm{^{-3}}}\leq T_{\rm{9}}\leq$10, which show the noticeable difference (up to $\sim$1.2 and $\sim$1.3 times) with those recommended in \cite{An99,NACREII} and \cite{An99,IIiadis2008}, respectively.

 \bigskip
  {\bf ACKNOWLEDGEMENT}
   \bigskip

    The authors are deeply grateful to L.D. Blokhintsev for   discussions and constructive suggestions and to S.V. Artemov for interest.   This work has been supported in part  by the Academy of Sciences   of the Republic of Uzbekistan  and by the Ministry of Education and Science of The Republic of Kazakhstan (grant No. AP05132062).
    
    \newpage

 \bigskip
  {\bf APPENDIX: THE BASED FORMULES OF THE MODIFIED $R$-MARTIX METHOD AND THE MODIFIED TWO-BODY POTENTIAL METHOD }
   \bigskip

Here we present only the idea and the main formulas for the astrophysical $S$ factors  of the modified $R$-method (see, for example,  Ref. \cite{Ar2012,BK1991} and references therein) and the MTBPM \cite{Igam07} specialized for the ${\rm{^9Be(}}p,\,\gamma{\rm{)^{10}B}}$ and ${\rm{^{16}O(}}p,\,\gamma{\rm{)^{17}F}}$ reactions, respectively,  as well as  the way of  obtaining  the ${\rm{^{19}F(}}p,\,\alpha{\rm{)^{16}O}}$ astrophysical $S$ factors  at thermonuclear energies, where  the  direct pole mechanism may be dominant.   The orbital angular momentum of the proton capture is  equal to 1  both for  the   ground and first three excited bound states  of the residual ${\rm{^{10}B}}$ nucleus and for    all the considered proton capture resonance states of ${\rm{^{10}B}}$ formed in the intermediate state.

The astrophysical $S$ factor is determined by
$$
S_{ij}(E)=Ee^{2\pi \eta}\sigma_{ij}(E), \eqno(A1)
$$where $\sigma_{ij}(E)$ is the reaction cross section, $E$  and $\eta$ are  the relative kinetic energy and the Coulomb parameter  of the colliding particles  $i$ and $j$. In (A1), the indexes at the astrophysical $S$ factor and the cross section denote mass numbers of the colliding particles.

According to \cite{Ar2012,BK1991}, within the framework of  the modified $R$-matrix method,  the total cross section for the     ${\rm{^9Be(}}p,\,\gamma{\rm{)^{10}B}}$ reaction populating the ground (${\rm{^{10}B_0}}$) and first three excited (${\rm{^{10}B}_f}$) bound states of the residual ${\rm{^{10}B}}$ nucleus  is given by
$$
 \sigma_{{\rm{1\,16}}}(E)=\sum_{f={\rm{0}}}^{{\rm{3}}}\sum_J\sigma_{J_{\rm{^{10}B}_f}\,J}(E). \eqno(A2)
$$Here $J$ and $J_{\rm{^{10}B}_f}$ are the total angular momentum of the colliding particles and the spin of the $f$th bound state in the ${\rm{^{10}B}}$   nucleus and
$$
\sigma_{J_{\rm{^{10}B}_f}\,J}(E)=\frac{\pi}{k^2}\frac{2J+1}{{\rm{8}}}\sum_{I\,l_i\,\lambda}\mid M_{J_{{\rm{^{10}B}_f}}\,J\,I\,l_i\,\lambda}(E)\mid, \eqno(A3)
$$where  $I$ and $l_i$ are the channel spin and the relative orbital momentum of the $p{\rm{^9Be}}$-scattering, respectively, $\lambda$ is the multipolarity order for the electromagnetic transition  and $k=\sqrt{{\rm{2}}\mu_{p{\rm{^9Be}}}E}/\hbar$ in which $\mu_{p{\rm{^9Be}}}$ is the reduced mass of $p$ and ${\rm{^9Be}}$.
By using this case, one notes that there is misprint  in the left-hand side of    Eq. (5) in  \cite{Ar2012}. There, the expression $\sigma_{J_i}$($E$)  should be replaced by that of  $\sigma_{J_f}$($E$).
 In  Eq. (A3), $M_{J_{{\rm{^{10}B}_f}}\,J\,I\,l_i\,\lambda}(E)$ is the  amplitude of the electromagnetic ($E\lambda$ and $M\lambda$) transition, which is represented in the form
$$
M_{J_{{\rm{^{10}B}_f}}\,J\,I\,l_i\,\lambda}(E)=M^{{\rm{(}}R_{j_{{\rm{0}}}};\,{\rm{(}}E\lambda,M\lambda{\rm{)}}{\rm{)}}}_{J_{{\rm{^{10}B}_f}}\,J\,I\,l_i\,\lambda}(E)+M^{{\rm{(DC}};\,E\lambda {\rm{)}}}_{J_{{\rm{^{10}B}_f}}\,J\,I\,l_i\,\lambda}(E)+ M^{{\rm{(DC}};\,M\lambda{\rm{)}}}_{J_{{\rm{^{10}B}_f}}\,J\,I\,l_i\,\lambda}(E),
\eqno(A4)
$$ where
$M^{{\rm{(}}R_{j_{{\rm{0}}}};\,{\rm{(}}E\lambda,M\lambda{\rm{)}}{\rm{)}}}_{J_{{\rm{^{10}B}_f}}\,J\,I\,l_i\,\lambda}$ is  the  proton capture amplitude of the $j_{{\rm{0}}}$th resonance state,  $M^{{\rm{(DC}};\,E\lambda{\rm{)}}}_{J_{{\rm{^{10}B}_f}}\,J\,I\,l_i\,\lambda}$ and    $M^{{\rm{(DC}};\,M\lambda{\rm{)}}}_{J_{{\rm{^{10}B}_f}}\,J\,I\,l_i\,\lambda}$ are  the direct proton $E\lambda$ and $M\lambda$  capture amplitudes in the $f$th bound state of ${\rm{^{10}B}}_f$, respectively.
In the single-level approximation, the $M^{{\rm{(}}R_{j_{{\rm{0}}}};\,{\rm{(}}E\lambda,M\lambda{\rm{)}}{\rm{)}}}_{J_{{\rm{^{10}B}_f}}\,J\,I\,l_i\,\lambda}$ amplitude can be represented in the form \cite{Holt1978,BK1991}
$$
 M^{{\rm{(}}R_{j_{{\rm{0}}}};\,{\rm{(}}E\lambda,M\lambda{\rm{)}}{\rm{)}}}_{J_{{\rm{^{10}B}_f}}\,J\,I\,l_i\,\lambda}=ie^{i(\sigma^{{\rm(c)}}_{l_i}-\delta^{{\rm{(HS)}}}_{l_i})}\frac{\left[\Gamma_{JIl_i}^p(E)\right]^{{\rm{1/2}}}
 \left[\Gamma_{J_{{\rm{^{10}B}_f}}\,J\,\lambda}^{\gamma}(E)\right]^{{\rm{1/2}}}}{E-E_{R_{j_{{\rm{0}}}}}+i\frac{\Gamma_{J}(E)}{{\rm{2}}}}.
\eqno(A5)
$$
Here $\sigma^{{\rm(c)}}_{l_i}$ and $\delta^{{\rm{(HS)}}}_{l_i}$ are the Coulomb and hard-sphere phase-shifts for the $p{\rm{^9Be}}$-scattering; $\Gamma_{JIl_i}^p(E)$ and $\Gamma_{J_{{\rm{^{10}B}_f}}\,J\,\lambda}^{\gamma}(E)$ are the partial protonic and radiative $\gamma$-withds for the resonant decays
 ${\rm{^{10}}B}_{j_{{\rm{0}}}}\to{\rm{^9Be}} +p$ and ${\rm{^{10}}B}_{j_{{\rm{0}}}}\to{\rm{^{10}B}_f}+\gamma$, respectively, and $\Gamma_{J}(E)$ is the total ($p$, $\alpha$ and $\gamma$) width. The energy dependence of the protonic and radiative $\gamma$-widths is given by the expressions
$$
\Gamma_{JIl_i}^p(E)=\frac{{\rm{2}}P_{l_i}(E){\rm{(}}\gamma_{JIl_i}^p{\rm{)}}^{{\rm{2}}}}{{\rm{1}}+{\rm{(}}\gamma _{JIl_i}^p{\rm{)}}^{{\rm{2}}}\left(\frac{dS_c}{dE}
 \right)_{E=E_{R_{j_{{\rm{0}}}}}}}
\eqno(A6a)
 $$and
 $$
 \Gamma_{J\,J_{{\rm{^{10}B}_f}}\,\lambda}^{\gamma}(E)=\frac{{\rm{2}}k_{\gamma}^{{\rm{2}}\lambda+{\rm{1}}}(E){\rm{(}}\gamma_{J\,J_{\rm{^{10}B}_f}\,\lambda}^{\gamma}{\rm{)}}^{{\rm{2}}}}
 {{\rm{1}}+{\rm{(}}\gamma_{J\,I\,l_i}^p{\rm{)}}^{{\rm{2}}}\left(\frac{dS_c}{dE}
 \right)_{E=E_{j_{{\rm{0}}}}}},
\eqno(A6b)
$$
where $k_{\gamma}$  is the photon momentum, $P_{l_i}$ is the penetrability factor, $S_c$ is the Thomas shift factor \cite{ThL1957}, and $\gamma_{...}^p$ and
$\gamma_{...}^{\gamma}$ are the partial  reduced protonic and radiative  $\gamma$-ray widths, respectively. The reduced  $\gamma_{J\,J_{\rm{^{10}B}_f}\,\lambda}^{\gamma}$ width involves the internal ($\gamma_{J\,J_{\rm{^{10}B}_f}\,\lambda}^{\gamma}$(int.))  and  external channel ($\gamma_{J\,J_{\rm{^{10}B}_f}\,\lambda}^{\gamma}$(ext.)) parts \cite{Holt1978}.  The external channel part  is a complex number and contains the ANC for ${\rm{^9Be}}+p\to{\rm{^{10}B}_f}$ as a factor and the channel radius ($r_{\rm{ch}}$) as a tree parameter.  The observable partial protonic and radiative $\gamma$-widths are given by
$$
\Gamma^p_{JIl_i}=|\Gamma^p_{JIl_i}{\rm{(}}E_{R_{j_{{\rm{0}}}}}{\rm{)}}|, \,\,\,\,
\Gamma_{J_{{\rm{^{10}B}_{j_{{\rm{0}}}}}}\,{\rm{^{10}B}}_f\,\lambda}^{\gamma}=|\Gamma_{J_{{\rm{^{10}B}_{j_{{\rm{0}}}}}}\,J_{{\rm{^{10}B}}_f}\,\lambda}^{\gamma}{\rm{(}}E_{R_{j_{{\rm{0}}}}}{\rm{)}}|.
\eqno(A7)
$$ One notes once more  that, in \cite{Satt1999},  the contribution of the external    part (so-called the proton-channel contribution) to the $\Gamma_{J_{{\rm{^{10}B}_{j_{{\rm{0}}}}}}\,J_{{\rm{^{10}B}}_f}\,\lambda}^{\gamma}$ width was ignored, which really contains the ANC's for ${\rm{^9Be}}+p\to{\rm{^{10}B}}_f$.

 The explicit expressions for  the direct capture amplitudes for the $E\lambda$ and $M\lambda$ transitions have rather cumbersome forms and, so, they are not presented here. Nevertheless,  we  note only  that, in the long wavelength  approximation,   they contain the radial integral, which has the form as
 $$
 I_{\lambda;J\,l_i}{\rm{(}}E{\rm{)}}=C_{{\rm{^{10}B}_f};\,J}\int_{r_{{\rm{ch}}}}^{\infty}drr^{\tilde{\lambda}}W_{-\eta_{{\rm{^{10}B}_f}};{\rm{3/2}}}{\rm{(2}}\kappa_{\rm{^{10}B}_f}
r{\rm{)(}}I_{l_i}{\rm{(}}kr{\rm{)}}-e^{i(\sigma^{{\rm(c)}}_{l_i}-\delta^{{\rm{(HS)}}}_{l_i})}O_{l_i}{\rm{(}}kr{\rm{))}}, \eqno(A8)
 $$ where $\tilde{\lambda}=\lambda$ and ($\lambda$-1) for the $E$1 and $M$1 transitions, respectively;
  $W_{-\eta_{{\rm{^{10}B}_f}};{\rm{3/2}}}$($\cdot\cdot\cdot$) is the Whittaker function;
     $\kappa_{\rm{^{10}B}_f}=\sqrt{{\rm{2}}\mu_{p{\rm{^9Be}}}\varepsilon_{\rm{^{10}B}_f}}/\hbar$ in which
     $\varepsilon_{\rm{^{10}B}_f}$  is the binding energy of the $f$th  bound state of  ${\rm{^{10}B}_f}$ in the (${\rm{^9Be}}+p$) channel, and $I_{l_i}{\rm{(}}kr{\rm{)}}$ and $O_{l_i}{\rm{(}}kr{\rm{)}}$ are the incoming and outcoming solutions of the radial Schr\"{o}dinger equation.

As is seen from the expression (A8), the powers of the total direct capture amplitude and the channel radiative $\gamma$-width are determined by the ANCs for ${\rm{^9Be}}+p\to{\rm{^{10}B}_f}$ ($f$=0--3). Hence,  introduction of information about the reliable ``indirect measured'' ANC's  to the resonance and direct capture amplitudes makes it possible to reduce   the uncertainty  of  the total $S_{{\rm{1\,16}}}$($E$) astrophysical $S$ factors  calculated for the ${\rm{^9Be}(}p,\,\gamma{\rm{)^{10}B}}$ reaction in thermonuclear energy region to a minimum, as it is possible.

 Within the framework of the MTBPM \cite{Igam07} (see, Ref. \cite{Art2009} also), the expression for the  astrophysical $S$ factors of the direct radiative capture  ${\rm{^{16}O}(}p,\,\gamma{\rm{)^{17}F}}$ reaction populating the ground  and first excited  states of the residual ${\rm{^{17}F}}$ nucleus   is presented in the form
$$
S_{l_{{\rm{^{17}F}}_j};\,{\rm{1\,16}}}{\rm{(}}E{\rm{)}}=C^{{\rm{2}}}_{{\rm{^{17}F}}_j;l_{{\rm{^{17}F}}_j}}{\cal{R}}_{l_{{\rm{^{17}F}}_j}}{\rm{(}}E;\,b_{{\rm{^{17}F}}_j}{\rm{)}}. \eqno(A9)
$$ Here $C_{{\rm{^{17}F}}_j;l_{{\rm{^{17}F}}_j}}$ is the ANC for ${\rm{^{16}O}}+p\to{\rm{^{17}F}_j}$ ($j$=1 and 2 for the ground and first excited states of  ${\rm{^{17}F}}$, respectively);  $b_{{\rm{^{17}F}}_j}$ is  the single-particle ANC, which determines the  amplitude of the ``tail'' of the radial component of the bound (${\rm{^{16}O}}+p$)  shell-model wave finction of the ${\rm{^{17}F}_j}$ nucleus calculated using
the Schr\"{o}dinger equation with the adopted Woods-Saxon potential, and ${\cal{R}}_{l_{{\rm{^{17}F}}_j}}{\rm{(}}E;\,b_{{\rm{^{17}F}}_j}{\rm{)}}=\sum_{\lambda}{\tilde{S}}_{l_{{\rm{^{17}F}}_j}\,\lambda;\,{\rm{1\,16}}}{\rm{(}}E;\,b_{{\rm{^{17}F}}_j}{\rm{)}}/b^{{\rm{2}}}_{{\rm{^{17}F}}_j}$ in which  ${\tilde{S}}_{l_{{\rm{^{17}F}}_j}\,\lambda;\,{\rm{1\,16}}}{\rm{(}}E;\,b_{{\rm{^{17}F}}_j}{\rm{)}}$ is the single-particle astrophysical $S$ factor \cite{An99}. As shown in  \cite{Art2009},  the free parameter  $b_{{\rm{^{17}F}}_j}$ in turn depends strongly from the geometric parameters (the radius $r_0$ and the diffuseness $a$) of the adopted Woods-Saxon potential, i.e., $b_{{\rm{^{17}F}}_j}=b_{{\rm{^{17}F}}_j}{\rm{(}}r_{{\rm{0}}},\,a{\rm{)}}$.
In \cite{Art2009}, the expression (A9) was used for determination of the ``indirect measured'' ANC's ($C_{{\rm{^{17}F}}_j}$) by means of replacement of the  astrophysical $S$ factor in the left hand side of Eq. (A9)  by  their experimental data  for each experimental point of the energy $E$ \cite{Morl1997}. This is  connected by the fact that the  reaction  for each the  fixed energy $E$ is strongly peripheral, since      the calculated values of the ${\cal{R}}_{l_{{\rm{^{17}F}}_j}}{\rm{(}}E;\,b_{{\rm{^{17}F}}_j}{\rm{)}}$ function  as a function of the tree   $b_{{\rm{^{17}F}}_j}{\rm{(}}r_{{\rm{0}}},\,a{\rm{)}}$ parameter   do not depend practically from variation of the free parameter.  The results  of  the square of the ANC's obtained in Ref. \cite{Art2009}   are also presented in Table \ref{table1}. Nevertheless, the ANC's obtained in \cite{Art2009} for each the experimental point of the energy $E$ have some spread associated with that of the data of Ref. \cite{Morl1997} plotted in Fig. \ref{fig5}.   On the other hand, the expression (A9) could  also  be used for calculation of $ S_{l_{{\rm{^{17}F}}_j}\,\lambda;\,{\rm{1\,16}}}{\rm{(}}E{\rm{)}}$  if the ``indirect measured'' ANC's above are known from other independent precisely measured  experimental data, e.g., from the experimental data for the peripheral  ${\rm{^{16}O(^3He}}, \,d{\rm{)^{17}F}}$ transfer proton  reaction. This issue is considered  in subsection B of Section VI.

We now show the way of applying  the expression (\ref{FBCS2}) for  obtaining the total cross sections (respective the astrophysical $S$ factor)
of the sub-barrier ${\rm{^{19}F(}}p,\,\alpha{\rm{)^{16}O}}$ triton transfer reaction considered in Section VI. For this end, we split the limit of changing for the  scattering  angle (0$\le\theta\le$180) in two   intervals: 0$\le\theta\le\theta_{{\rm{max}}}$, where  good agreement between the experimental DCS's  and the calculated ones occurs, and $\theta_{{\rm{max}}}<\theta\le$180$^o$, where there is the noticeable discrepancy   between the experimental DCS's  and the calculated ones  (see Fig. \ref{fig3}).  Then, from Eq. (\ref{FBCS2}) one has
$$
\sigma_{{\rm{1\,19}}}{\rm{(}}E{\rm{)}}=\sigma_<{\rm{(}}E{\rm{)}}+\sigma_>{\rm{(}}E{\rm{)}}, \eqno(A10a)
$$
$$
\sigma_<{\rm{(}}E{\rm{)}}=
{\rm{2}}\pi C_{\alpha}^{{\rm{2}}}C_{{\rm{^{19}F}}_j}^{{\rm{2}}}\int_{{\rm{0}}}^{\theta_{{\rm{max}}}}
d\theta\sin\theta\tilde{\sigma}_{r_{{\rm{0}}}}{\rm{(}}j_{{\rm{^{19}F}}};E,\,\theta{\rm{)}}, \eqno(A10b)
$$
$$
\sigma_>{\rm{(}}E{\rm{)}}={\rm{2}}\pi\int_{\theta_{{\rm{max}}}}^{{\rm{180}}}d\theta\sin\theta\frac{d\sigma}{d\Omega}, \eqno(A10c)
$$where $j_{{\rm{^{19}F}}}$= 1/2, $E$=$E_i$, $r_{{\rm{0}}}$= 1.25 fm, and the $C_{\alpha}^{{\rm{2}}}$ and $C_{{\rm{^{19}F}}_j}^{{\rm{2}}}$ ANC's are known.  For each considered fixed proton projectile  energy,   the  integral in the right-hand side of Eq. (A10b) can be taken numerically using the calculated $\tilde{\sigma}_{r_{{\rm{0}}}}{\rm{(}}j_{{\rm{^{19}F}}};E,\,\theta{\rm{)}}$ function and the corresponding $C_{{\rm{^{19}F}}_j}^{{\rm{2}}}$ given  in Table  \ref{table1}.
 Due to the fact that the  experimental angular distributions of the  ${\rm{^{19}F(}}p,\,\alpha{\rm{)^{16}O}}$  reaction plotted in Fig. \ref{fig3}${\rm{(}}a{\rm{)}}$-- ${\rm{(}}f{\rm{)}}$ monotonic decease (including for $\theta>\theta_{{\rm{max}}}$), similarly as it is done in Ref. \cite{Raim1990},    the Legendre polynomial expansion
$$
\left(\frac{d\sigma}{d\Omega}\right)_{\theta>\theta_{{\rm{max}}}}=\sum_{n={{\rm 0}}}^{{\rm{2}}}a_n{\rm{(}}E{\rm{)}}P_l{\rm{(}}\cos\theta{\rm{)}}
\eqno(A11)
$$
 for the integrand function
of the integral (A10c)  is applied   for reproducing   the corresponding experimental DCS's in the angular range of $\theta>\theta_{{\rm{max}}}$. The values of the fitted $a_n{\rm{(}}E{\rm{)}}$ coefficients of the expression (A11) providing well description of the experimental data in the corresponding angular range   are given in Table 2. They   can   be  used for calculating of the integral (A10c). The results of calculations of ($d\sigma/d\Omega$)$_{\theta>\theta_{{\rm{max}}}}$ are displayed   in Fig. \ref{fig3} by the dotted curves. Thus, Eqs. (A10)--(A11) allow us to calculate the total cross sections  (respective the astrophysical $S$ factors) for the ${\rm{^{19}F(}}p,\,\alpha{\rm{)^{16}O}}$ reaction  for each fixied center-of-mass  projectile energy from the EXP-1978 and EXP-2015 data.

 \newpage

 \newpage

\newpage

\newpage

\newpage

\newpage

  {\selectlanguage{english}
\begin{table}[t]
\begin{center}
 \caption{\label{table1}{\small  Reaction,  energy $E_x$,  set of the optical potentials (set), virtual decay $B\to\,A\,+\,a$,  orbital and total angular momentums ($l_B$, $j_B$),   square modulus of    the nuclear vertex constant $|G_B|^{{\rm {2}}}$( $G_B=G_{Aa;l_Bj_B}$) for the virtual decay   $B\to \,A\,+\,a$ and   the corresponding  ANC  $C_B^{{\rm {2}}}$ ($C_B=C_{Aa;l_Bj_B}$) $\,A\,+\,a\,\to\,B$. Figures in brackets are experimental and theoretical uncertainty, respectively, whereas those in square brackets are weighed mean derived from   the ANC's (NVC's) values  for  the sets 1 and 2. }}
\vspace{2mm}
{\footnotesize
\begin{tabular}{lllllll}
 \hline
\multicolumn{1}{c}{$A$($x$, $y$)$B$}
 &\multicolumn{1}{c}{\,\,\,  $E_x$, MeV}
  &\multicolumn{1}{c}{ set}
 &\multicolumn{1}{c}{ $B\to\,A\,+\,a$}
&\multicolumn{1}{c}{\,\,$l_B$, $j_B$}
&\multicolumn{1}{c}{$|G_B|^{{\rm {2}}}$, fm}
&\multicolumn{1}{c}{$C_B^{{\rm {2}}}$, fm$^{{\rm{-1}}}$}\\
 \hline\\
\,\,\,\,\,\,\,\,\,\,\,\,\,\,\,\,\,1&\,\,\,\,\,\,2&3&\,\,\,\,\,\,\,\,\,\,\,\,\,\,\,\,\,4&\,\,\,\,\,\,\,\,5&\,\,\,\,\,\,\,\,\,\,\,\,\,\,\,\,\,6&\,\,\,\,\,\,\,\,\,\,\,\,\,\,\,\,\,7\\
  \hline\\
 ${\rm {^9Be(^{10}B,^9Be)^{10}B_0}}$&
    100 \cite{Mukh2}& 1& ${\rm{^{10}B_0}}\to\, {\rm{^9Be}} \,+p$&1, 3/2&  0.72($\pm$0.06;$\pm$0.02)&4.22($\pm$0.33;$\pm$0.10)\\
    &  & & & &0.72 $\pm$0.06&4.22$\pm$0.35\\
    &    & &  & &  0.84$\pm$0.07 \cite{Mukh2} &4.91$\pm$0.39 \cite{Mukh2}\\
&  & 2& & &0.77($\pm$0.06;$\pm$0.02)&4.49($\pm$0.37;$\pm$0.11)\\
  &  & & & &0.77 $\pm$0.07&4.49$\pm$0.39\\
     &  & &  & &  0.92$\pm$0.07 \cite{Mukh2} &5.35$\pm$0.42 \cite{Mukh2}\\
    &  & 1+2& & &[0.75($\pm$0.04;$\pm$0.02)]&[4.35($\pm$0.24;$\pm$0.14)]\\
     &  & & & &[0.75$\pm$0.05]&[4.35$\pm$0.28]\\
     &  & &  & &[0.87$\pm$0.08] \cite{Mukh2}&[5.06$\pm$0.46] \cite{Mukh2}\\
     ${\rm{^9Be(^3He}}, \,d{\rm)^{10}B_0}$          &22.3--32.5 \cite{Ar96}  & &  & &0.63--0.73 \cite{Ar96}&3.67--4.20 \cite{Ar96}\\
     ${\rm {^9Be(^{10}B,^9Be)^{10}B_1}}$&  100 \cite{Mukh2}& 1& ${\rm{^{10}B_1}}\to\, {\rm{^9Be}} \,+p$
     &1, 1/2&  0.22($\pm$0.02;$\pm$0.01)&1.31($\pm$0.10;$\pm$0.03)\\
    &  & & & &0.22 $\pm$0.02&1.31$\pm$0.11\\
    &    & &  & &  0.21$\pm$0.03 \cite{Mukh2} &1.23$\pm$0.17 \cite{Mukh2}\\
    && 2&
     &&  0.25($\pm$0.02;$\pm$0.01)&1.47($\pm$0.17;$\pm$0.03)\\
    &  & & & &0.25 $\pm$0.02&1.47$\pm$0.17\\
    &    & &  & &  0.23$\pm$0.03 \cite{Mukh2} &1.34$\pm$0.19 \cite{Mukh2}\\
    &  & 1+2& & &[0.23($\pm$0.02;$\pm$0.01)]&[1.33($\pm$0.09;$\pm$0.03)]\\
    &  & & & &[0.23$\pm$0.02]&[1.33$\pm$0.10]\\
     &  & &  & &[0.22$\pm$0.04] \cite{Mukh2}&[1.27$\pm$0.21] \cite{Mukh2}\\
     ${\rm{^9Be(^3He}}, \,d{\rm)^{10}B_1}$          &22.3--32.5 \cite{Ar96}  & &  & &0.62--0.69 \cite{Ar96}&3.16--4.02 \cite{Ar96}\\
    && 1&&1, 3/2&  0.60($\pm$0.05;$\pm$0.01)&3.53($\pm$0.27;$\pm$0.09)\\
    &  & & & &0.60 $\pm$0.05&3.53$\pm$0.29\\
    &    & &  & &  0.57$\pm$0.05 \cite{Mukh2} &3.33$\pm$0.29 \cite{Mukh2}\\
&  & 2& & &0.68($\pm$0.05;$\pm$0.02)&3.98($\pm$0.31;$\pm$0.09)\\
  &  & & & &0.68 $\pm$0.06&3.98$\pm$0.32\\
   &  & &  & &  0.23$\pm$0.05 \cite{Mukh2} &3.63$\pm$0.32 \cite{Mukh2}\\
  &  & 1+2& & &[0.64($\pm$0.04;$\pm$0.04)]&[3.74($\pm$0.22;$\pm$0.23)]\\
    &  & & & &[0.64$\pm$0.05]&[3.74$\pm$0.32]\\
     &  & &  & &[0.59$\pm$0.07] \cite{Mukh2}&[3.43$\pm$0.42] \cite{Mukh2}\\
          ${\rm{^9Be(^3He}}, \,d{\rm)^{10}B_1}$          &22.3--32.5 \cite{Ar96}  & &  & &0.37--0.42 \cite{Ar96}&2.16--2.45 \cite{Ar96}\\
     ${\rm {^9Be(^{10}B,^9Be)^{10}B_2}}$&
    100 \cite{Mukh2}& 1& ${\rm{^{10}B_2}}\to\, {\rm{^9Be}} \,+p$&1, 3/2&  0.58($\pm$0.04;$\pm$0.01)&3.39($\pm$0.26;$\pm$0.08)\\
    &  & & & &0.58 $\pm$0.05&3.39$\pm$0.27\\
   &    & &  & &  0.72$\pm$0.08 \cite{Mukh2} &4.22$\pm$0.44 \cite{Mukh2}\\
&  & 2& & &0.66($\pm$0.05;$\pm$0.02)&3.88($\pm$0.30;$\pm$0.10)\\
  &  & & & &0.77 $\pm$0.05&3.88$\pm$0.32\\
   &  & &  & &  0.79$\pm$0.08 \cite{Mukh2} &4.60$\pm$0.48 \cite{Mukh2}\\
  &  & 1+2& & &[0.61($\pm$0.04;$\pm$0.02)]&[3.60($\pm$0.24;$\pm$0.24)]\\
    &  & & & &[0.61$\pm$0.06]&[3.60$\pm$0.34]\\
     &  & &  & &[0.74$\pm$0.09] \cite{Mukh2}&[4.35$\pm$0.59] \cite{Mukh2}\\
                 ${\rm{^9Be(^3He}}, \,d{\rm)^{10}B_2}$          &32.5 \cite{Ar96}  & &  & &1.25$\pm$0.14 \cite{Ar96}&7.29$\pm$0.82 \cite{Ar96}\\
      \end{tabular}}
\end{center}
\end{table}

\newpage

{\selectlanguage{english}
\begin{table}[t]
\begin{center}
 \caption{{continuation of Table 1}}
\vspace{2mm}
{\footnotesize
\begin{tabular}{lllllll}
\hline\\
\,\,\,\,\,\,\,\,\,\,\,\,\,\,\,\,\,1&\,\,\,\,\,\,2&3&\,\,\,\,\,\,\,\,\,\,\,\,\,\,\,\,\,4&\,\,\,\,\,\,\,\,5&\,\,\,\,\,\,\,\,\,\,\,\,\,\,\,\,\,6&\,\,\,\,\,\,\,\,\,\,\,\,\,\,\,\,\,7\\
  \hline\\
     ${\rm {^9Be(^{10}B,^9Be)^{10}B_3}}$&    100 \cite{Mukh2}
    & 1& ${\rm{^{10}B_3}}\to\, {\rm{^9Be}} \,+p$&1, 1/2&
      0.055($\pm$0.004;$\pm$0.001)&0.32($\pm$0.03;$\pm$0.01)\\
    &  & & & &0.055 $\pm$0.004&0.32$\pm$0.03\\
    &    & &  & &  0.048$\pm$0.009 \cite{Mukh2} &0.28$\pm$0.05 \cite{Mukh2}\\
&  & 2& & &0.046($\pm$0.004;$\pm$0.002)&0.27($\pm$0.02;$\pm$0.01)\\
  &  & & & &0.046 $\pm$0.005&0.27$\pm$0.03\\
   &  & &  & &  0.0513$\pm$0.0085 \cite{Mukh2} &0.30$\pm$0.05 \cite{Mukh2}\\
 &  & 1+2& & &[0.055($\pm$0.003;$\pm$0.005)]&[0.32($\pm$0.02;$\pm$0.03)]\\
    &  & & & &[0.055$\pm$0.005]&[0.32$\pm$0.03]\\
     &  & &  & &[0.050$\pm$0.010] \cite{Mukh2}&[0.29$\pm$0.06] \cite{Mukh2}\\
   & &1& &1, 3/2&
      0.155($\pm$0.012;$\pm$0.004)&0.91($\pm$0.07;$\pm$0.02)\\
    &  & & & &0.155 $\pm$0.012&0.91$\pm$0.07\\
    &    & &  & &  0.14$\pm$0.02 \cite{Mukh2} &0.80$\pm$0.10 \cite{Mukh2}\\
&&2   & &&0.13($\pm$0.01;$\pm$0.01)&0.76($\pm$0.07;$\pm$0.04)\\
  &  & & & &0.13 $\pm$0.01&0.76$\pm$0.08\\
   &  & &  & &  0.15$\pm$0.02 \cite{Mukh2} &0.87$\pm$0.11 \cite{Mukh2}\\
  &  & 1+2& & &[0.15($\pm$0.01;$\pm$0.01)]&[0.89($\pm$0.06;$\pm$0.08)]\\
    &  & & & &[0.15$\pm$0.02]&[0.89$\pm$0.10]\\
     &  & &  & &[0.14$\pm$0.02] \cite{Mukh2}&[0.82$\pm$0.12] \cite{Mukh2}\\
                 ${\rm{^9Be(^3He}}, \,d{\rm)^{10}B_3}$          &32.5 \cite{Ar96}  & &  & &0.26$\pm$0.03  \cite{Ar96}&1.52$\pm$0.17 \cite{Ar96}\\
       ${\rm{^{16}O(^3He}}, d{\rm {)^{17}F_0}}$ &29.75 \cite{Mukh1999}& 1 &${\rm{^{17}F}}\to\, {\rm{^{16}O}} \,+p$ &2, 5/2&0.179($\pm$0.018;$\pm$0.009)&1.14($\pm$0.12;$\pm$0.06)\\
    &    & &  & &  0.179$\pm$0.020   &1.14$\pm$0.13  \\
    &    & &  & &  0.16 \cite{Mukh1999} &1.0 \cite{Mukh1999}\\
     & & 2 & & &0.206($\pm$0.021;$\pm$0.010)&1.31($\pm$0.14;$\pm$0.07)\\
      &    & &  & &  0.206$\pm$0.024   &1.31$\pm$0.15  \\
       &    & &  & &  0.18 \cite{Mukh1999}   &1.10  \cite{Mukh1999} \\
  &  &1+2 & & &[0.190($\pm$0.014;$\pm$0.013)]&[1.21($\pm$0.09;$\pm$0.08)]\\
     &    & &  & & [0.190$\pm$0.019]   &[1.21$\pm$0.12]  \\
     &    & &  & &  [0.170$\pm$0.016] \cite{Mukh1999}   &[1.08$\pm$0.10] \cite{Mukh1999} \\
 &18;34 \cite{Ar96}     & &  & & 0.16  \cite{Ar96}   &1.02 \cite{Ar96}\\
                ${\rm{^{16}O(}}p, \gamma{\rm {)^{17}F_0}}$ &     & &  & &  0.17$\pm$0.02 \cite{Art2009}   &1.09$\pm$0.11  \cite{Art2009} \\
   The $p{\rm{^{16}O}}$-scattering:             &     & &  & &  0.12$\,{\rm{^{a}}}$ \cite{Blokh2018}   &0.77${\,\rm{^{a}}}$  \cite{Blokh2018} \\
    the phase-shifts             &     & &  & &  0.37${\,\rm{^{b}}}$ \cite{Blokh2018}   &5.48${\,\rm{^{b}}}$ \cite{Blokh2018} \\
  analysis   &     & &  & &  0.17${\,\rm{^{c}}}$ \cite{YarBaye2011}   &1.09${\,\rm{^{c}}}$ \cite{YarBaye2011} \\
     ${\rm{^{16}O(^3He}}, d{\rm {)^{17}F_1}}$  &29.75 \cite{Mukh1999}& 1 &${\rm{^{17}F}}\to\, {\rm{^{16}O}} \,+p$ &0, 1/2&916($\pm$96;$\pm$46)&5840($\pm$611;$\pm$292)\\
     &    & &  & &  916$\pm$106  &5840$\pm$667 \\
      &    & &  & &  939 \cite{Mukh1999} &5980 \cite{Mukh1999}\\
     & & 2 & & &1053($\pm$110;$\pm$53)&6713($\pm$703;$\pm$335)\\
      &    & &  & & 1053$\pm$122&6713$\pm$779\\
           &    & &  & & 1099 \cite{Mukh1999} &7000 \cite{Mukh1999}\\
            &    &1+2 &  & & [975($\pm$72;$\pm$68)]&[6216($\pm$461;$\pm$432)]\\
                      &    &  &  & & [975$\pm$99]   &[6216$\pm$632]  \\
        &    & &  & &  [1019$\pm$107] \cite{Mukh1999}   &[6490$\pm$680] \cite{Mukh1999} \\
          &18; 34  \cite{Ar96}    & &  & & 840; 819 \cite{Ar96}   &5355; 5122 \cite{Ar96}\\
${\rm{^{16}O(}}p, \gamma{\rm {)^{17}F_1}}$ &     & &  & &  893$\pm$35 \cite{Art2009}   &5700$\pm$225  \cite{Art2009} \\
 The $p{\rm{^{16}O}}$-scattering:             &     & &  & &  1629${\,\rm{^{a}}}$\cite{Blokh2018}   &10384${\,\rm{^{a}}}$  \cite{Blokh2018} \\
    the phase-shifts             &     & &  & &  1245${\,\rm{^{b}}}$ \cite{Blokh2018}   &7939${\,\rm{^{b}}}$ \cite{Blokh2018} \\
  analysis &  & &  & &  893${\,\rm{^{c}}}$ \cite{YarBaye2011}   &5700${\,\rm{^{c}}}$ \cite{YarBaye2011} \\
  \end{tabular}
 }
\end{center}
   \end{table}

  \newpage

  \newpage

{\selectlanguage{english}
\begin{table}[t]
\begin{center}
 \caption{{continuation of Table 1}}
\vspace{2mm}
{\footnotesize
\begin{tabular}{lllllll}
\hline\\
\,\,\,\,\,\,\,\,\,\,\,\,\,\,\,\,\,1&\,\,\,\,\,\,2&3&\,\,\,\,\,\,\,\,\,\,\,\,\,\,\,\,\,4&\,\,\,\,\,\,\,\,5&\,\,\,\,\,\,\,\,\,\,\,\,\,\,\,\,\,6&\,\,\,\,\,\,\,\,\,\,\,\,\,\,\,\,\,7\\
  \hline\
      ${\rm{^{19}F(}}p,\,\alpha {\rm {)^{16}O}}$  &0.250  \cite{HL1978}&  &${\rm{^{19}F}}\to\, {\rm{^{16}O}} \,+\,t$ &0, 1/2&13.5($\pm$2.1;$\pm$0.7)&618.1($\pm$95.2;$\pm$30.9)\\
 EXP-1978  &0.350   &  & & &13.2($\pm$1.4;$\pm$0.7)&605.0($\pm$63.4;$\pm$30.3)\\
 &0.450   &  & & &11.9($\pm$1.3;$\pm$0.6)&544.8($\pm$60.6;$\pm$27.2)\\
  weighed mean &
  &  & &  &12.7($\pm$0.9;$\pm$0.5)&583.5($\pm$39.8;$\pm$23.3)\\
  & &  &  &  &12.7$\pm$1.0&583.5$\pm$46.1\\
 EXP-2015  &0.327  \cite{Ivano2015}&  &  & &28.1($\pm$2.7;$\pm$1.4)&1290.3($\pm$124.0;$\pm$64.3)\\
   &0.387   &  & & &29.2($\pm$3.2;$\pm$1.5)&1341.3($\pm$144.7;$\pm$66.6)\\
 &0.486   &  & & &27.2($\pm$2.9;$\pm$1.4)&1248.1($\pm$134.6;$\pm$62.5)\\
 weighed mean &
  &  & &  &28.1($\pm$1.7;$\pm$0.8)&1291.1($\pm$77.2;$\pm$37.2)\\
  & &  &  &  &28.1$\pm$1.9&1291.1$\pm$85.7\\
   \hline
   \end{tabular}}
 \end{center}
  ${\rm{^{a}}}$The effective-range function (RRF) method.\\
   ${\rm{^{b}}}$The $\Delta$ method.\\
  ${\rm{^{c}}}$The effective-range  expansion method for the $p{{\rm{^{16}O}}}$-scattering function.
  \end{table}

\newpage

\begin{table}
 \caption{\label{table3}  The values of the fitted coefficients $a_n{\rm{(}}E{\rm{)}}$x10$^{{\rm{4}}}$ of the expession (A11) of Appendix for different center-of-mass  energies $E$. $E$ in keV and $a_n{\rm{(}}E{\rm{)}}$ in mb/sr.}
 \bigskip
 \begin{tabular}{llll} \hline
 \multicolumn{1}{c}{ $E$}& \multicolumn{1}{c}{  $a_{{\rm{0}}}{\rm{(}}E{\rm{)}}$}& \multicolumn{1}{c}{  $a_{{\rm{1}}}{\rm{(}}E{\rm{)}}$}& \multicolumn{1}{c}{  $a_{{\rm{2}}}{\rm{(}}E{\rm{)}}$} \\
 \hline
237.5&0.4079&0.0524&-0.1803\\
310.7&8.9528&6.8962&\,\,0.7359\\
332.5&3.5218&-2.7426&-5.5391\\
367.7&18.60&12.80&-2.6297\\
427.5&26.80&16.80&-4.8594\\
461.7&78.20&40.20&-36.40\\
\hline
 \end{tabular}
 \end{table}

\newpage

{\selectlanguage{english}
\begin{table}[t]
\begin{center}
 \caption{\label{table2}The fitted resonant parameters for the astrophysical  $S$ factor of the ${\rm{^9Be(}}p,\gamma{\rm{)^{10}B}}$ reaction.}
\vspace{2mm}
{\footnotesize
\begin{tabular}{llllll}
 \hline
 \hline
Resonance & Compilation & Zahnow, & Wulf,  & A.Sattorov,  & Present\\
parameters & \cite{Ajzen1988}  &{\it et. al.} \cite{Zah1995}&{\it et. al.} \cite{Wulf1998} &{\it et. al.} \cite{Satt1999} & work \\ \hline
$J^{\pi}$ & 1$^{-}$  & 1$^{-}$ & 1$^{-}$& 1$^{-}$ & 1$^{-}$ \\
$E_{{\rm{1}}}$ [keV] & 287$\pm$5  &342$\pm$27  & 295  & 296 & 282 \\
$\Gamma_{{\rm{1}}}$ [keV] & 120$\pm$5  &297$\pm$27 & 145 & 140 & 140 \\
${\Gamma^{p}}/{\Gamma_{1}}$ & 0.3  &  &0.3& 0.35 & 0.35 \\
$\Gamma^{\gamma}$ [eV] & 4.8 & 4.8&1.8 & 1.2$^{\,\rm{a}}$ & 1.2 \\
\hline
$J^{\pi}$ & 2$^{\,{\rm{^+}}{\rm{b}}}$  & 2$^{+}$ & 2$^{+}$& 2$^{+}$ & 2$^{+}$ \\
$E_{2}$ [keV] & $892 \pm 2$  & 890$\pm$1.8 & 890&890 & 890 \\
$\Gamma_{{\rm{2}}}$ [keV] & 72$\pm$4 & 81.0$\pm$2.7 & 79.2 & 80 & 83 \\
$\Gamma^{p}/\Gamma_{{\rm{2}}}$ & $\approx$0.65 & & & 0.75 & 0.75 \\
$\Gamma^{\gamma}$ [eV] & 25.8 &  & &25.8$^{\,\rm{c}}$ & 25.8 \\
 \hline
$J^{\pi}$ & 0$^{+}$  & 0$^{+}$ & 0$^{+}$& 0$^{+}$ & 0$^{+}$ \\
$E_{{\rm{3}}}$ [keV] & 972 & 972 & 972& 972  & 957 \\
$\Gamma_{{\rm{3}}}$ [keV] &2.65$\pm$0.18  &  & & & 2.7 \\
$\Gamma^{p}/\Gamma_{{\rm{3}}}$ & 1.0& 1.0  & 1.0 & 1.0 & 1.0 \\
$\Gamma^{\gamma}$ [eV] & 8.5 &  8.5  & 8.5& 8.5$^{\,\rm{c}}$ & 6.5 \\
\hline
$J^{\pi}$ & 2$^{-}$  & 2$^{-}$ & 2$^{-}$ & 2$^{-}$ & 2$^{-}$ \\
$E_{{\rm{4}}}$ [keV] & 1161  & 1265& 1215 & 1196 & 1196 \\
$\Gamma_{{\rm{4}}}$ [keV] & 210$\pm$60 & 387$\pm$27&190 & 290 & 290 \\
$\Gamma^{p}/\Gamma_{{\rm{4}}}$ [eV] & $\approx$0.65) &  &0.72& 0.52 & 0.52 \\
$\Gamma^{\gamma}$ [eV] & 8.5 &  &5.8& 7.9$^{\,\rm{c}}$ & 7.9 \\
     \hline
      \end{tabular}}
\end{center}
${\rm{^a}}$This parameter was taken from Ref. \cite{AM1975}.\\
${\rm{^b}}$See table 10.16 in Ref. \cite{Ajzen1988} and p.p. 5 and 6 of Ref. \cite{Satt1999}.\\
${\rm{^c}}$This parameter was taken from Ref. \cite{Ajzen1988}.\\
\end{table}

\newpage

\begin{table}
 \caption{\label{table4}  Rates $N_A\langle\sigma_{ij}v_{ij}\rangle$ of the  ${\rm{^9Be(}}p,\gamma {\rm{)^{10}B}}$ and ${\rm{^{16}O(}}p,\gamma {\rm{)^{17}F}}$ reactions in the dependence from  the temperature $T_{{\rm{9}}}$ (K) in the unit of 10$^{\rm{9}}$.}
 \bigskip
 \begin{tabular}{llllllll} \hline
 \multicolumn{8}{c}{$N_A\langle\sigma_{ij}v_{ij}\rangle$, cm$^{{\rm{3}}}$mol${\rm{^{-1}}}$s${\rm{^{-1}}}$}\\
 \hline
 $T_{{\rm{9}}}$, &\multicolumn{3}{c}{ ${\rm{^9Be(}}p,\gamma {\rm{)^{10}B}}$ }
  &$T_{{\rm{9}}}$  &\multicolumn{3}{c}{ ${\rm{^{16}O(}}p,\gamma {\rm{)^{17}F}}$ }
  \\
K &our work&\cite{An99}(ratio)&\cite{NACREII}(ratio)& K&our work &\cite{An99}(ratio)&\cite{IIiadis2008}(ratio)\\
  \hline
 \,\,\,  1&\,\,\,\,2   &\,\,\,\, 3&\,\,\,\, \,\,4&\, 5&\,\,\,\,\,\, 6&\,\,\,\,\,\, 7&\,\,\,\,\,\, 8\\
  \hline
0.010&3.81[-13]&3.87[-13](0.98)&4.29[-13](0.89)&0.010&  8.58[-25]&6.73[-25](1.27)&7.20[-25](1.19)\\
0.011&1.61[-12]&1.64[-12](0.98)&1.81[-12](0.89)&0.011&7.97[-24]&7.10[-24](1.12)&7.58[-24](1.05)\\
0.012&5.74[-12]&5.90[-12](0.97)&6.47[-12](0.87)&0.012&6.13[-23]&5.71[-23](1.07)&6.08[-23](1.01)\\
0.013&1.79[-11]&1.85[-11](0.97)&2.02[-11](0.89)&0.013&3.99[-22]&3.68[-22](1.09)&3.91[-22](1.02)\\
0.014&4.99[-11]&5.18[-11](0.96)&5.64[-11](0.88)&0.014&2.20[-21]&1.97[-21](1.12)&2.09[-21](1.05)\\
0.015&1.27[-10]&1.32[-10](0.96)&1.43[-10](0.87)&0.015&1.03[-20]&9.06[-21](1.14)&9.60[-21](1.08)\\
0.016&2.97[-10]&3.10[-10](0.96)&3.35[-10](0.89)&0.016&4.20[-20]&3.65[-20](1.15)&3.87[-20](1.09)\\
0.018&1.34[-9]&1.41[-9](0.95)&1.51[-9](0.89)&0.018&4.92[-19]&4.30[-19](1.14)&4.54[-19](1.08)\\
0.020&4.89[-9]&5.17[-9](0.95)&5.50[-9](0.89)&0.020&4.05[-18]&3.59[-18](1.13)&3.77[-18](1.07)\\
0.025&6.53[-8]&6.99[-8](0.93)&7.32[-8](0.89)&0.025&2.78[-16]&2.50[-16](1.11)&2.62[-16](1.06)\\
0.030&4.70[-7]&5.08[-7](0.93)&5.24[-7](0.90)&0.030&6.96[-15]&6.32[-15](1.10)&6.59[-15](1.06)\\
0.040&8.29[-6]&9.11[-6](0.91)&9.15[-6](0.91)&0.040&7.49[-13]&6.89[-13](1.09)&7.16[-13](1.05)\\
0.050&6.36[-5]&7.05[-5](0.90)&6.94[-5](0.92)&0.050&2.06[-11]&1.91[-11](1.08)&1.98[-11](1.04)\\
0.060&3.01[-4]&3.34[-4](0.90)&3.24[-4](0.93)&0.060&2.57[-10]&2.39[-10](1.07)&2.48[-10](1.03)\\
0.070&1.00[-3]&1.15[-3](0.90)&1.11[-3](0.93)&0.070&1.91[-9]&1.79[-9](1.07)&1.86[-9](1.03)\\
0.080&2.89[-3]&3.18[-3](0.91)&3.05[-3](0.95)&0.080&9.96[-9]&9.37[-9](1.06)&9.71[-9](1.03)\\
0.090&6.86[-3]&7.50[-3](0.91)&7.16[-3](0.96)&0.090&4.01[-8]&3.78[-8](1.06)&3.92[-8](1.02)\\
0.10&1.45[-2]&1.57[-2](0.92)&1.50[-2](0.96)&0.10&1.33[-7]&1.26[-7](1.05)&1.30[-7](1.02)\\
0.11&2.78[-2]&2.99[-2](0.93)&2.85[-2](0.97)&0.11&3.76[-7]&3.57[-7](1.05)&3.70[-7](1.02)\\
0.12&4.9[-2]&5.31[-2](0.93)&5.06[-2](0.98)&0.12&9.45[-7]&9.01[-7](1.05)&9.31[-7](1.01)\\
\hline
   \end{tabular}
 \end{table}

 \newpage

 \begin{table}
 \caption{\label{table6} Continued Table 4}
 \bigskip
 \begin{tabular}{llllllll} \hline
 \multicolumn{8}{c}{$N_A\langle\sigma_{ij}v_{ij}\rangle$, cm$^{{\rm{3}}}$mol${\rm{^{-1}}}$s${\rm{^{-1}}}$}\\
 \hline
  \,\,\,  1&\,\,\,\,2   &\,\,\,\, 3&\,\,\,\, \,\,4&\, 5&\,\,\,\,\,\, 6&\,\,\,\,\,\, 7&\,\,\,\,\,\, 8\\
  \hline
0.13&8.35[-2]&8.88[-2](0.94)&8.47[-2](0.99)&0.13&2.15[-6]&2.06[-6](1.05)&2.12[-6](1.01)\\
0.14&1.34[-1]&1.41[-1](0.95)&1.35[-1](0.99)&0.14&4.51[-6]&4.32[-6](1.05)&4.46[-6](1.01)\\
0.15&2.05[-1]&2.16[-1](0.95)&2.06[-1](1.00)&0.15&8.83[-6]&8.48[-6](1.04)&8.74[-6](1.01)\\
0.16&3.05[-1]&3.20[-1](0.95)&3.05[-1](1.00)&0.16&1.63[-5]&1.57[-5](1.04)&1.62[-5](1.01)\\
0.18&6.16[-1]&6.42[-1](0.96)&6.13[-1](1.01)&0.18&4.81[-5]&4.65[-5](1.03)&4.78[-5](1.01)\\
0.20&1.13&1.18(0.97)&1.13(1.01)&0.20&1.22[-4]&1.18[-4](1.03)&1.21[-4](1.00)\\
0.25&3.99&4.08(0.98)&3.93(1.01)&0.25&7.78[-4]&7.57[-4](1.03)&7.77[-4](1.00)\\
0.30&1.06[1]&1.07[1](0.99)&1.04[1](1.02)&0.30&3.17[-3]&3.09[-3](1.02)&3.17[-3](0.99)\\
0.35&2.30[1]&2.31[1](0.99)&2.25[1](1.02)&0.35&9.67[-3]&9.44[-3](1.02)&9.67[-3](1.00)\\
0.40&4.32[1]&4.30[1](1.00)&4.22[1](1.02)&0.40&2.41[-2]&2.36[-2](1.02)&2.43[-2](0.99)\\
0.45&7.25[1]&7.16[1](1.01)&7.07[1](1.03)&0.45&5.22[-2]&5.09[-2](1.03)&5.25[-2](0.99)\\
0.50&1.11[2]&1.09[2](1.02)&1.08[2](1.03)&0.50&1.01[-1]&9.84[-2](1.03)&1.01[-1](0.99)\\
0.60&2.17[2]&2.11[2](1.03)&2.11[2](1.03)&0.60&2.99[-1]&2.90[-1](1.03)&3.01[-1](0.99)\\
0.70&3.54[2]&3.44[2](1.03)&3.45[2](1.03)&0.70&7.07[-1]&6.83[-1](1.03)&7.12[-1](0.99)\\
0.80&5.18[2]&5.04[2](1.03)&5.08[2](1.02)&0.80&1.43&1.38(1.04)&1.44(0.99)\\
0.90&7.11[2]&6.95[2](1.02)&6.98[2](1.02)&0.90&2.59&2.49(1.04)&2.61(0.99)\\
1.00&9.41[2]&9.20[2](1.02)&9.24[2](1.02)&1.00&4.31&4.13(1.04)&4.34(0.99)\\
1.25&1.76[3]&1.70[3](1.03)&1.71[3](1.03)&1.25&1.19[1]&1.13[1](1.05)&1.19[1](0.99)\\
1.50&3.08[3]&2.92[3](1.05)&2.93[3](1.05)&1.50&2.55[1]&2.42[1](1.05)&2.56[1](1.00)\\
1.75&4.98[3]&4.61[3](1.08)&4.62[3](1.08)&1.75&4.68[1]&4.40[1](1.06)&4.68[1](1.00)\\
2.00&7.39[3]&6.75[3](1.09)&6.73[3](1.10)&2.00&7.69[1]&7.19[1](1.07)&7.67[1](1.00)\\
2.50&1.31[4]&1.16[4](1.13)&1.16[3](1.13)&2.50&1.67[2]&1.54[2](1.08)&1.66[2](1.01)\\
3.00&1.89[4]&1.64[4](1.16)&1.65[4](1.15)&3.00&3.00[2]&2.73[2](1.10)&\\
3.50&2.42[4]&2.07[4](1.17)&2.09[4](1.16)&3.50&4.77[2]&4.30[2](1.11)&\\
4.00&2.86[4]&2.46[4](1.16)&2.45[4](1.17)&4.00&6.98[2]&6.24[2](1.12)&\\
5.00&3.45[4]&2.90[4](1.19)&2.94[4](1.17)&5.00&1.27[3]&1.12[3](1.13)&\\
6.00&3.76[4]&3.15[4](1.19)&3.18[4](1.18)&6.00&1.99[3]&1.75[3](1.14)&\\
7.00&3.88[4]&3.29[4](1.18)&3.27[4](1.19)&7.00&2.85[3]&2.49[3](1.15)&\\
8.00&3.89[4]&3.26[4](1.19)&3.26[4](1.19)&8.00&3.83[3]&3.34[3](1.15)&\\
9.00&3.84[4]&3.23[4](1.19)&3.20[4](1.20)&9.00&4.90[3]&4.27[3](1.15)&\\
10.00&3.76[4]&3.24[4](1.16)&3.10[4](1.21)&10.00&6.06[3]&5.27[3](1.15)&\\
  \hline
   \end{tabular}
 \end{table}


\newpage


\begin{thebibliography} {*}
\bibitem{Fowler1984}
W.A. Fowler, Rev. Mod. Phys. {\bf 56}, 149 (1984).
\bibitem{Rolf88}
 C. Rolfs and W.S. Rodney, Cauldrons in the Cosmos,(University of
Chicago Press,Chicago and London 1988).
\bibitem{Adel2011}
E.G. Adelberger, A.Garcia, R. G. Hamish Robertson, K.A. Snover,
A.B.Balantekin, K. Heeger, M. J, Ramsey-Musolf, D. Bemmerer, A.
Junghans, C.A. Bertulani, J.-W. Chen, H. Costantini, P. Prati, M.
Couder, E. Uberseder, M. Wiescher, R. Cyburt, B. Davids, S. J.
Freedman, M. Gai, D. Gazit, L. Gialanella, G. Imbriani, U. Greife,
M. Hass, W.C. Haxton, T. Itahashi, K. Kubodera, K. Langanke, D.
Leitner, M. Leitner, P. Vetter, L. Winslow, L.E. Marcucci, T.
Motobayashi, A.M. Mukhamedzhanov, R.E. Tribble, Konneth M. Nollett,
F.M. Nunes, T.-S. Park, P.D. Parker, R. Schiavilla, E.C. Simpson, C.
Spitaleri, F. Strieder, H.-P. Trautvetter, K. Suemmerer, S. Typel,
Rev. Mod. Phys. {\bf 83}, 195 (2011).
\bibitem{Blokh2010}
L. D. Blokhintsev, R. Yarmukhamedov, S.V. Artemov, I. Boztosun, S.B. Igamov, Q.I. Tursunmakhtov, and M.K. Ubaydullaeva, Uzb. J. Phys. {\bf 12},  217 (2010).
\bibitem{Yarm2013}
R. Yarmukhamedov, and  Q.I. Tursunmahatov, {\it The  Universe Evolution: Astrophysical and nuclear aspects. Edit. I. Strakovsky and L. D. Blokhintsev}.
(New York, NOVA publishers, 2013), pp.219--270.
\bibitem{Trib2014}
R.E. Tribble, C.A. Bertulani, M. La Cognata, A.M. Mukhamedzhanov, and C. Spitaleri, Rep. Prog. Phys. {\bf 77}, 901 (2014).
\bibitem{Satt1999}
A. Sattarov, A.M. Mukhamedzhanov, A. Azhari, C.A. Gagliardi, L. Trache, and R.E. Tribble, Phys. Rev. C {\bf 60}, 035801 (1999).
 \bibitem{Art2009}
S.V. Artemov, S.B. Igamov, Q.I. Tursunmakhatov, and R. Yamukhamedov,
Bull. RAN. Ser. Phys. {\bf 73}, 176 (2009) [Izv. RAN. Ser. Fiz. {\bf 73}, 165 (2009)].
\bibitem{Ivano2015}
I. Lombardo, D. Dell'Aquila, A.Di Leva, I. Indelicato, M. La Cognata, M. La Commara, A. Ordine, V. Rigato, M. Romoli, F. Rosalo, G. Spadaccini, C. Spitareli, A. Tumino, and M. Viliance, Phys. Lett. B{\bf 778}, 178 (2015).
\bibitem{BD1998}
 D. Baye, P. Descouvemont, and M. Hesse, Phys. Rev. C{\bf 58} (1998) 545.
\bibitem{Wulf1998}
E.A. Wulf, M.A. Godwin, J.F. Guillemette, C.M. Laymon, R.M. Prier, B.J. Rice, V. Spraker, D.R. Tilley, H.R. Weller, Phys. Rev. C{\bf 58}, (1998) 517.
 \bibitem{Blokh2018}
 L.D. Blokhintsev, A.S. Kadyrov, A.M. Mukhamedzhanov, D.A. Savin, Phys. Rev. C {\bf 98}, 064610 (2018).
\bibitem{HAS1991}
H. Herndl, H. Abele, G. Staudt,  B. Bach, K. Gr\"{u}n, H. Scsribany, H. Oberhummer, G. Raimann, Phys. Rev. C{\bf 44}, R952 (1991).
\bibitem{Ceciletal1992}
F.E. Cecil, D. Ferg, H. Liu, J.C. J.C. Scorby, J.M. McNeil, and P.D. Kunz, Nucl. Phys. A{\bf 539}, 75 (1992.
\bibitem{Zah1995}
D. Zahnow, C Angulo, M. Junker, C. Rofs, S. Schimidt, W.H. Schulte, and E. Somorjai,  Nucl. Phys. A{\bf 589}, 95 (1995).
\bibitem{Rolfs1973}
C. Rolfs, Nucl. Phys. A{\bf 217}, 29 (1973).
\bibitem{Mukh2}
 A. M. Mukhamedzhanov, H. L. Clark, C. A. Gagliardi, Y.-W. Lui, L. Thache,
R. E. Tribble, H. M. Xu, X. G. Zho\'{u}, V. Burjan, J. Cejpek, V. Kroha, and F. Carstoiu,   Phys. Rev.
C{\bf 56}, 1302 (1997).
\bibitem{Mukh10}
 Sh.S. Kajumov, A.M. Mukhamedzhanov, R.
 Yarmukhamedov, and I. Borbely, Z. Phys. A {\bf 336}. 297 (1990).
\bibitem{YTur2019}
R. Yarmukhamedov and K.I. Tursunmakhatov, submitted to Phys. Rev. C, 2019.
\bibitem{Holt1978}
R.J. Holt, H.E. Jackson, R.M. Laszewski, J.E. Monahan, and J.R. Spechi, Phys. Rev. C{\bf 18}, 1962 (1978).
\bibitem{An99}
 C. Angulo, M. Arnould, M. Rayet, P. Descouvemont, D. Baye,
 C. Leclercq-Willain, A. Coc, S. Barhoumi, P. Aguer, C. Rolfs, R. Kunz, J. W. Hammer,
    A. Mayer, T.  Paradellis, S. Kossionides, C. Chronidou, K. Spyrou,
    S. Degl'Innocenti, G.  Fiorentini, B. Ricci, S. Zavatarelli, C. Providencia,
    H. Wolters, J. Soares, C. Grama, J. Rahighi, A. Shotter, M. L. Rachti, Nucl. Phys.
    A{\bf 656}, 3 (1999).
\bibitem{Morl1997}
R. Morlock, R. Kunz,  A. Mayer, M. Jaeger, A. M\"{u}ller, J.M. Hummer,  P. Mohr, H. Oberhummer, G. Staudt, and V. K\"{o}lle,  Phys. Rev. Lett. {\bf 79}, 3837 (1997).
\bibitem{Mukh1999}
C.A. Gagliardi, R.E. Tribble, A. Azhari,    H. L. Clark,  Y.-W. Lui, A. M. Mukhamedzhanov, A. Sattarov, L. Thache, V. Burjan, J. Cejpek, V. Kroha, \v{S}. \v{P}iskov, and J. Vincour, Phys. Rev. C {\bf 59}, 1149 (1999).
\bibitem{Ar96}
S. V. Artemov, I. R. Gulamov, E. A. Zaparov, I. Yu. Zotov, and G. K. Nie,
Yad. Fiz. {\bf 59}, 454 (1996)[Phys. At. Nucl. {\bf 59}, 428 (1996)].
\bibitem{YarBaye2011}
R. Yarmukhamedov and D. Baye, Phys. Rev. C{\bf 84}, 24603 (2011).
\bibitem{HL1978}
H. Lorenz-Wirzha, Ph.D. thesis, Universit\"{a}t, M\"{u}ster, 1978.
 \bibitem{HAS1991}
H. Herndl, H. Abele, G. Staudt,  B. Bach, K. Gr\"{u}n, H. Scsribany, H. Oberhummer, G. Raimann, Phys. Rev. C {\bf 44}, R952 (1991).
\bibitem{Blok77}
L.D. Blokhintsev, I. Borbely, E.I. Dolinskii, Phys. Part. Nucl. {\bf 8}, 485 (1977).
  \bibitem{YaBl2018}
R. Yarmukhamedov and L. D. Blokhintsev, Phys. At. Nucl. {\bf 81}, 616 (2018).
\bibitem{Pl1973}
G. R. Plattner, R. D. Viollier, D. Trautmann, K. Alder, Nucl. Phys. A {\bf 206}, 513 (1973).
\bibitem{Art2008}
S.V. Artemov, A.B. Bajajin, S.B. Igamov, G.K. Nie, and R. Yamukhamedov,
Yad. Fiz. {\bf 71}, 998 (2008)[Phys. At. Nucl. {\bf 71}, 1025 (2008)].
\bibitem{Mukh1995}
A.M. Mukhamedzhanov, R.E. Tribble, and N.K. Timofeyuk, Phys. Rev. C{\bf 51}, 3472 (1995).
\bibitem{Ar2012}
 S. V. Artemov, S. B. Igamov, Q. I. Tursunmakhatov, and R. Yarmukhamedov, Phys. Atom. Nucl. {\bf 75}, 291 (2012).
\bibitem{Igam07}
S. B. Igamov, and R. Yarmukhamedov, Nucl. Phys. A {\bf 781}, 247
(2007).
\bibitem{Ajzen1988}
 F. Ajzenberg-Selove, Nucl. Phys. A {\bf 490}, 1 (1988).
 \bibitem{AM1975}
 W. Auw\"{a}rter and V. Mayer, Nucl. Phys. A{\bf 242}, 129 (1975).
  \bibitem{Con2015}
M. La Cognata, C. Palmirini,  C. Spitareli, I. Indelicato, A.M. Mukhamedzhanov,  I. Lombardo, and O. Trippella, Astron. J. {\bf 805}, 128 (2015).
\bibitem{Isoya1958}
A. Isoya, H. Ohmura, and T. Momota, Nucl. Phys. {\bf 7}, 116, (1958).
\bibitem{RBF1996}
R.B. Firestone, {\it Table of Isotopes CD-ROM, Info nucl A=1--20} (Version 1.0, March 1996), p. 158.
 \bibitem{NACREII}
 Y. Xu, K. Takahashi, S. Goriely, M. Arnould, M. Ohta, and H. Utsunomiya, Nucl. Phys. A{\bf 918}, 61 (2013).
  \bibitem{IIiadis2008}
  C. IIiadis, C. Angulo, P. Discouvemont, M. Lugaro, and P. Mohr, Phys. Rev. C {\bf 77}, 045802 (2008).
 \bibitem{BK1991}
 F.C. Barker and T. Kajino, Aust. J. Phys. {\bf 44}, 369 (1991).
  \bibitem{ThL1957}
A.M. Lane and R.G. Thomas, Rev. Mod. Phys. {\bf 30}, 257 (1957).
\bibitem{Raim1990}
G. Raimann, B. Bach, K. Gr\"{u}n, H. Herndl, H. Oberhummer, S. Engstler, C. Rolfs, H. Abele, R. Neu, and
G.  Staudt, Phys. Lett. {\bf B249}, 191 (1990).


\newpage


  \end{thebibliography}
\end{document}